\documentclass[aps,twocolumn, longbibliography,a4paper,superscriptaddress,pra]{revtex4-1}

\usepackage{amsfonts}
\usepackage{amsmath}
\usepackage{amssymb}
\usepackage{hyperref}
\usepackage{graphicx}
\usepackage{changes}
\usepackage{soul}

\newcommand{\inp}{\mathrm{in}}
\newcommand{\out}{\mathrm{out}}

\newcommand{\sd}{\mathcal{S}}
\newcommand{\ssd}{\bar{\mathcal{S}}}

\newcommand{\modeda}{ \delta a}
\newcommand{\modedadag}{ \delta a^\dagger}

\newcommand{\dain}{\delta \hat{a} _\inp}
\newcommand{\daout}{\delta \hat{a} _\out}

\newcommand*{\Vpi}{\ensuremath{V_\pi}}
\newcommand*{\MD}{\ensuremath{\mathrm{MD}}}
\newcommand*{\unit}[1]{\ensuremath{\,\mathrm{#1}}}
\newcommand*{\s}[1]{\ensuremath{_\mathrm{#1}}}	

\begin{document}
	
\title{Cryogenic electro-optic interconnect for superconducting devices}

\author{Amir Youssefi}\thanks{These authors contributed equally.}
\author{Itay Shomroni}\thanks{These authors contributed equally.}
\affiliation{Institute of Physics, \'Ecole Polytechnique F\'ed\'erale de Lausanne (EPFL), CH-1015 Lausanne, Switzerland}
\author{Yash J. Joshi}
\affiliation{Institute of Physics, \'Ecole Polytechnique F\'ed\'erale de Lausanne (EPFL), CH-1015 Lausanne, Switzerland}
\affiliation{Indian Institute of Science Education and Research, Pune 411008, India}
\author{Nathan Bernier}
\author{Anton Lukashchuk}
\author{Philipp Uhrich}
\author{Liu Qiu}
\author{Tobias~J.~Kippenberg}
\email[]{tobias.kippenberg@epfl.ch}
\affiliation{Institute of Physics, \'Ecole Polytechnique F\'ed\'erale de Lausanne (EPFL), CH-1015 Lausanne, Switzerland}

\maketitle

\textbf{
	Encoding information onto optical fields using electro-optical modulation is the backbone of modern telecommunication networks, offering vast bandwidth and low-loss transport via optical fibers~\cite{winzer2018fiber}. 
	For these reasons, optical fibers are also replacing electrical cables for short range communications within data centers~\cite{kachris2012survey}.
	Compared to electrical coaxial cables, optical fibers also introduce two orders of magnitude smaller heat load from room to milli-Kelvin temperatures, making optical interconnects based on electro-optical modulation an attractive candidate for interfacing superconducting quantum circuits~\cite{devoret_superconducting_2013,martinis_quantum_2020,blais_quantum_2020} and hybrid superconducting devices~\cite{clerk_hybrid_2020}.
	Yet, little is known about optical modulation at cryogenic temperatures.
	Here we demonstrate a proof-of-principle cryogenic electro-optical interconnect,
	showing that currently employed Ti-doped lithium niobate phase modulators~\cite{wooten2000review} are compatible with operation down to $\mathbf{800\,mK}$---below the typical operation temperature of conventional microwave amplifiers based on high electron mobility transistors (HEMTs)~\cite{pospieszalski1988fets, duh1988ultra}---and maintain their room temperature Pockels coefficient.
	We utilize cryogenic electro-optical modulation to perform
	spectroscopy of a superconducting circuit optomechanical system, measuring optomechanically induced transparency (OMIT)~\cite{Weis2010, Zhou2013,safavi2011electromagnetically,TeufelStrongCoupling}.
	\begin{figure}[ht!]
		\includegraphics[scale=1]{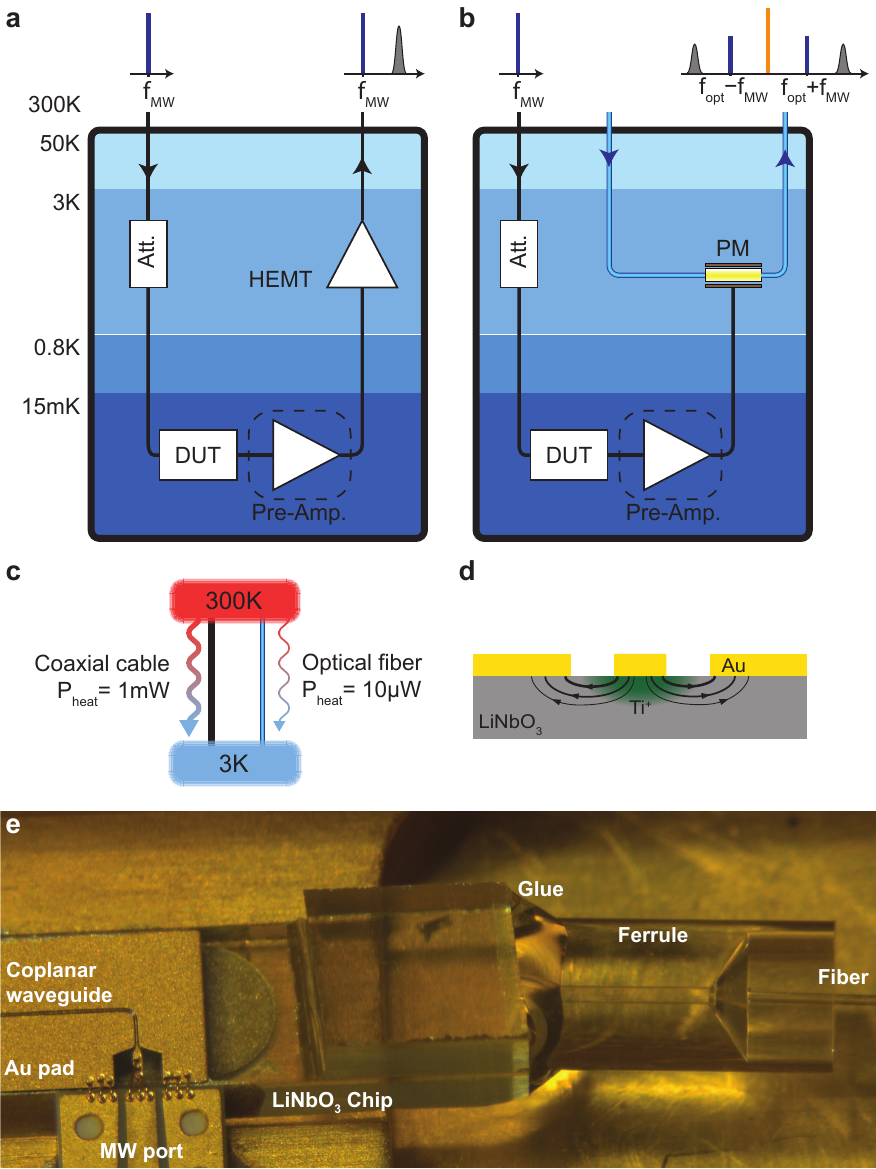}
		\caption{\footnotesize \linespread{1}
			\textbf{Principle of a cryogenic electro-optical interconnect for readout of superconducting devices.}
			\textbf{a}, Simplified schematic of a conventional readout of a device under test (DUT) in a dilution fridge using a cryogenic HEMT amplifier.
			The dashed box indicates an optional quantum-limited pre-amplifier not used in this work. The devices are interrogated by input microwave signals that are attenuated to reduce thermal noise, and amplified using an HEMT amplifier at $3\unit{K}$.
			\textbf{b},	Principle of a cryogenic electro-optic readout scheme using an electro-optical phase modulator. The DUT is interrogated using the same microwave input line, but the microwave signals are converted to the optical domain at 3~K, reducing thermal load.
			\textbf{c}, Conducted heat through a typical cryogenic coaxial cable and optical fiber, between room temperature and $3\unit{K}$.
			\textbf{d}, Schematic cross-section of a $z$-cut LiNbO$_3$ electro-optic phase modulator.
			\textbf{e}, Microscope photo of the commercial phase modulator used in the experiment, showing the coupling region between fiber and LiNbO$_3$ chip.
			\label{fig:1}}
	\end{figure}
	In addition, we encode thermomechanical sidebands from the microwave domain onto an optical signal processed at room temperature. 
	Although the currently achieved noise figure is significantly higher 
	than that of a typical HEMT,
	substantial noise reduction should be attainable by
	harnessing recent advances in integrated modulators~\cite{Loncar2018,he2019,thiele2020,chakraborty2020cryogenic}, by increasing the modulator length, or by using materials with a higher electro-optic coefficient~\cite{Abel2019, Abel_BTO_cryo},
	leading to noise levels on par with HEMTs.
	Our work highlights the potential of
	electro-optical
	modulators for massively parallel readout for emerging quantum computing~\cite{devoret_superconducting_2013, martinis_quantum_2020, blais_quantum_2020,clerk_hybrid_2020} or cryogenic classical computing~\cite{braginski2019} platforms.}

Optical modulators are ubiquitous in our information society and encode electrical signals in optical carriers that can be transported over fiber.
Initially only used for long-haul communications, optical fiber links are now also replacing electrical cables for short range communications within data centers~\cite{Cheng2018, Thomson2016, blumenthal2000all}.
This is motivated by the high power consumption of electrical interconnects that spurred the development of optical interconnects based on silicon photonics~\cite{Thomson2016}.
Such interconnects may also be used in the future for on-board chip-to-chip communication~\cite{sun2015single, miller2000optical}.

A similar challenge is foreseeable in superconducting quantum circuits, where recent advances~\cite{devoret_superconducting_2013, martinis_quantum_2020, blais_quantum_2020} have highlighted the potential associated with scaling superconducting qubit technology~\cite{arute2019quantum}.
Currently, significant efforts are underway to scale the number of qubits~\cite{krinner2019}.
As a result, one of the challenges that future progress in superconducting circuits will face is to massively increase the number of microwave control and readout lines while preserving the base temperature and protecting qubits from thermal noise.

Figure~\ref{fig:1}a shows a prototypical measurement chain of a single superconducting device-under-test (DUT) that operates at the $15\unit{mK}$ stage of a dilution refrigerator.
Coaxial cables are used to transmit output signals to the room temperature as well as to send control signals to the cold stages of the fridge.
To read out GHz microwave signals, 
a high electron mobility transistor (HEMT) amplifier with low-added-noise [$n\s{add}\sim 10\unit{quanta/(s\cdot Hz)}$] is typically employed that operates at the $3\unit{K}$ stage and amplifies the DUT output signal for further processing outside the cryostat.
Although HEMTs are not quantum-limited~\cite{pospieszalski1988fets, duh1988ultra}, the development of Josephson junction-based pre-amplifiers~\cite{yamamoto2008flux, macklin2015near, siddiqi2004rf, castellanos2008amplification} that operate at the $15\unit{mK}$ stage have allowed near-quantum-limited microwave amplification.
%

The presence of coaxial cables introduces additional heat load from room temperature into the cold stages of the refrigerator, which poses significant barrier to the scalability of such systems~\cite{krinner2019}.
In contrast, optical fibers have superior thermal insulation, reducing the heat load per line by two orders of magnitude (Fig.~\ref{fig:1}c).
Optical fibers additionally exhibit ultralow signal losses, $\sim\! 0.2\unit{dB/km}$, compared to $\sim\!3\unit{dB/m}$ at GHz frequencies for coaxial lines (compensated by the HEMT amplification).
%
Note also that thermal noise is completely negligible at optical frequencies.
Optical fibers could therefore provide a solution to scaling the number of lines 
without the concomitant heating.
For this approach, a critical component are transducers that convert input microwave signals to the optical domain, which are compatible with low temperature operation and are sufficiently efficient to ensure low noise conversion of microwave to optical signals.
Indeed, substantial efforts are underway to create quantum-coherent interfaces between the microwave and optical domains~\cite{lauk2020perspectives}.
To date, quantum coherent conversion schemes based on piezo-electromechanical~\cite{jiang2019lithium, jiang2019efficient}, magneto-optical~\cite{bartholomew2019chip}, and optomechanical~\cite{Higginbotham2018, forsch2020microwave, arnold2020converting, andrews2014bidirectional,chu2020perspective} coupling have been  developed.
In addition, schemes based on cavity electro-optics~\cite{tsang2010} have been demonstrated using bulk~\cite{rueda2016efficient,Rueda2019,hease2020} or integrated~\cite{Faneaar2018,mckenna2020,holzgrafe2020} microwave cavities coupled via the Pockels effect to an optical cavity mode.
Yet, all these schemes have in common that they transduce \emph{narrowband} microwave signals to the optical domain.
While this ability is critical for future quantum networks, an optical replacement for the currently employed HEMT amplifiers may be required for scaling control lines.
One route is therefore to use \emph{broadband} optical modulators as already used today in telecommunication networks.
While this approach may yield lower conversion efficiency compared to systems employing narrowband resonant cavities, continued improvements in design, and new material systems, can render it competitive, especially given its relative simplicity.

Here we explore this potential and replace the HEMT amplifier with a LiNbO$_3$-based optical phase modulator (PM), the workhorse of modulator technology, in order to directly transduce the DUT microwave output signal onto sidebands around the optical carrier field (Fig.~\ref{fig:1}b), detectable using standard homodyne or heterodyne detection schemes at ambient temperatures.
%
To illustrate the principle of the readout, we consider the operating principle of a PM.
Optical PMs are based on the Pockels effect (Fig.~\ref{fig:1}d) and induce a phase shift on the input optical field $E\s{in}(t)$, proportional to the voltage $V(t)$ applied on the input microwave port of the device,
\begin{equation}
E\s{out}(t) = E\s{in}(t) e^{-i\pi V(t)/\Vpi} \approx E\s{in}(t)[1-i\pi V(t)/\Vpi],
\end{equation}
where the half-wave voltage $\Vpi$ is the voltage at which the phase shift is $\pi$, and typical $V(t)\ll\Vpi$ is assumed.
The relation between microwave (field operator $\hat{b}$) and optical (field operator $\hat{a}$) photon flux spectral densities~\cite{clerk2010}, $\ssd_{bb}$ and $\ssd_{aa}$ respectively, can be written as (see Appendix)
\begin{equation}
\ssd_{aa}[\omega\s{opt}\pm\omega\s{MW}] = G\times (\ssd_{bb}[\omega\s{MW}] + n\s{add})
\label{eq:input_output_flux}
\end{equation}
where $\omega\s{MW}$ and $\omega\s{opt}$ are the microwave signal and optical carrier frequencies, $n\s{add}$ is the added noise of the transducer (referred to the input), and the transduction gain $G$ is the number of transduced optical photons per microwave input photon, given by (see Appendix):
\begin{equation}
G = P\s{opt}\frac{\omega\s{MW}}{\omega\s{opt}}\frac{\pi^2 Z_0}{2\Vpi^2}
\label{eq:G}
\end{equation}
where $P\s{opt}$ is the power of the optical carrier at the output of the PM, and $Z_0$ its input microwave impedance.
In this experiment, we employ a commercial (Thorlabs LN65S-FC), $z$-cut traveling wave Ti-doped LiNbO$_3$ PM with specified bandwidth of $10\unit{GHz}$ and $\Vpi=7.5\unit{V}$ at $10\unit{GHz}$ (Fig.~\ref{fig:1}e).
We use a $1555\unit{nm}$ fiber laser as the optical source.
The typical incident optical power on the PM is $15\unit{mW}$.
The optical transmission of the PM was reduced during the first cooldown, and measured at 23\%.
Additional details on the cryogenic optical setup are given in the Appendix.

Previous works investigated the temperature dependence of the electro-optic coefficient and refractive index of congruent LiNbO$_3$ at low frequencies down to $7\unit{K}$~\cite{Herzog2008}.
Commercial $x$-cut LN modulators were also tested down to $10\unit{K}$, showing a slight change in $\Vpi$ 
from its room temperature value~\cite{Morse1994,McConaghy1996}.
Ref.~\cite{Yoshida1999} discusses the behavior of LiNbO$_3$ modulators with superconducting electrodes down to $4\unit{K}$.

To date, however, such modulators have not been used in a dilution refrigerator to directly read out a superconducting device.

\begin{figure*}[ht!]
	\includegraphics[width=\textwidth]{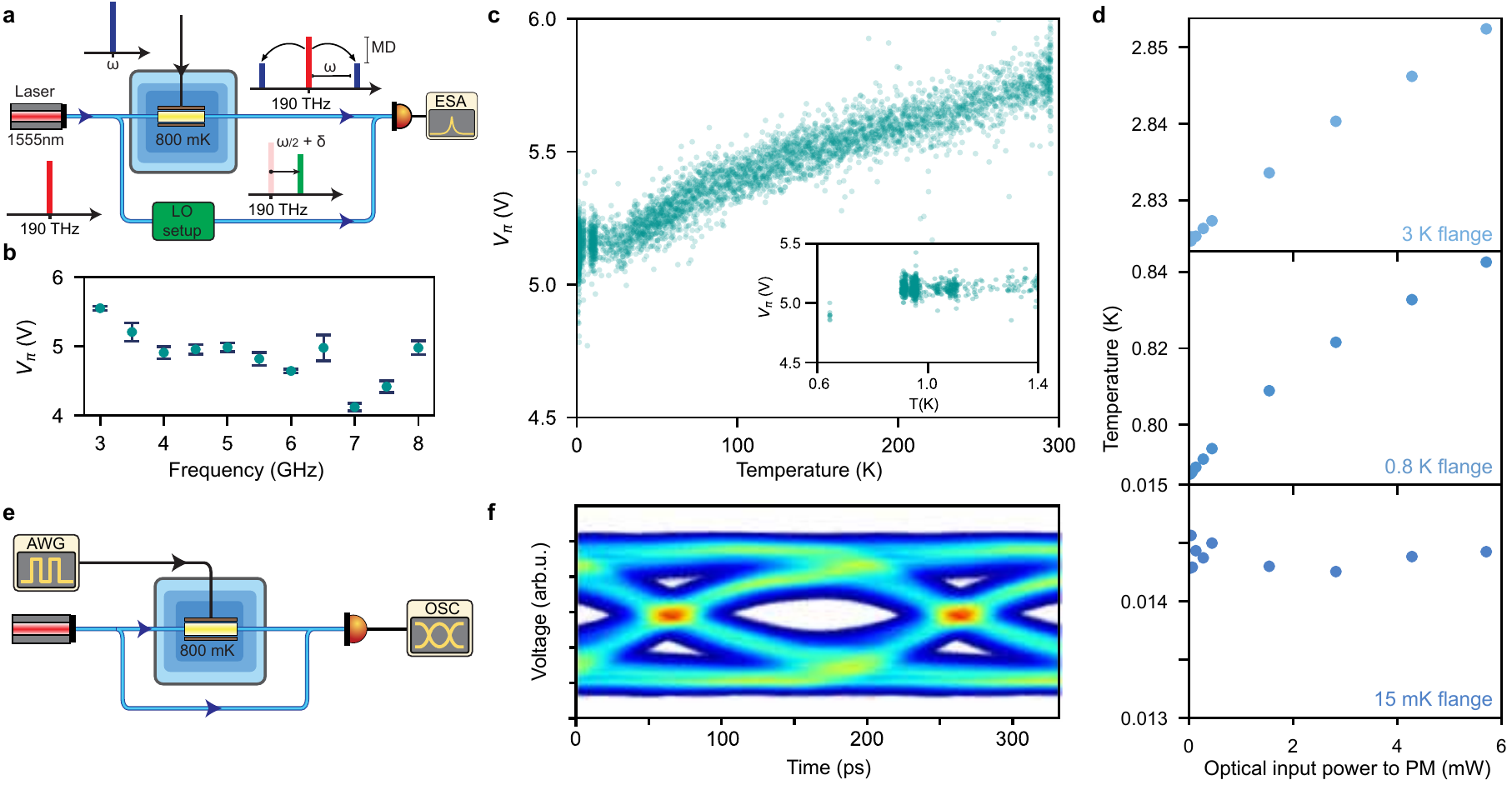}
	\caption{
		\footnotesize \linespread{1}
		\textbf{Cryogenic characterization of a LiNbO$_3$ phase modulator.}
		\textbf{a},
		Experimental setup for low temperature characterization of the phase modulator.
		\textbf{b},
		plot of $\Vpi$ vs.~frequency at $800\unit{mK}$.
		\textbf{c},
		Characterization of $\Vpi$ at 5\unit{GHz} vs.~temperature from room temperature to $800\unit{mK}$.
		\textbf{d},
		Measurement of heating due to optical dissipation when the phase modulator is mounted on the $800\unit{mK}$ flange.
		Plot of the steady state temperature vs.~input laser power of the $3\unit{K}$, $800\unit{mK}$, and $15\unit{mK}$ flanges.
		\textbf{e},
		Experimental setup for phase shift keying detection.
		RF signal from waveform generator is directly applied on a phase modulator.
		After homodyne detection the electrical signal is recorded on a fast oscilloscope.
		\textbf{f},
		Eye-diagram of an optical signal phase-modulated at a rate of $5\unit{GBaud}$, the bit error ratio is $5\times 10^{-5}$.
	}
	\label{fig:2}
\end{figure*}

\subsection*{Characterization}%
To characterize the electro-optic behavior of the device at cryogenic temperatures, we mount the PM on the $800\unit{mK}$ flange of the dilution fridge. 
We directly drive the microwave port of the PM at frequency $\omega\s{MW}$ using a microwave source outside the fridge, generating sidebands around the optical carrier frequency (Fig.~\ref{fig:2}a).
The half-wave voltage $\Vpi$ is determined from the modulation depth $\MD$, defined as the ratio of the power in one of sidebands to the power in the carrier,
\begin{equation}
\Vpi = \pi\sqrt{\frac{Z_0 P\s{MW}}{2\,\MD}},
\label{eq:vpi}
\end{equation}
where $P\s{MW}$
is the power at the microwave input port of the PM.
We measure $\MD$ by beating the output optical signal with a local oscillator (LO) with frequency $\omega\s{opt}+\omega\s{MW}/2+\delta$, 
generating two closely-spaced beatnotes at $\omega\s{MW}/2\pm\delta$, due to the carrier and the high-frequency sideband.
Using Eq.~\eqref{eq:vpi} we extract $\Vpi$ by sweeping the microwave power and measuring $\MD$.
Figure~\ref{fig:2}c shows $\Vpi$ at 5\unit{GHz} monitored as the fridge is cooled down from room temperature to $800\unit{mK}$,
and Fig.~\ref{fig:2}b shows $\Vpi$ at different frequencies at $800\unit{mK}$.
Importantly, $\Vpi$ does not change substantially from the room temperature value.

To investigate the effect of heating caused by optical dissipation in the PM, we measured the steady state temperature of different flanges of the dilution fridge when the PM is mounted on the $800\unit{mK}$ flange.
The results are shown in Fig.~\ref{fig:2}d.
In the Appendix, by comparing to a calibrated heater, we show that these temperature increases can be attributed to optical power loss within the PM package (and not, e.g, light leakage into the fridge volume).
This allows quantitative comparison with, e.g., heat dissipation due to a HEMT, and suggests reduced heat load in high optical transmission devices.

\begin{figure*}[ht]
	\includegraphics[width=\textwidth]{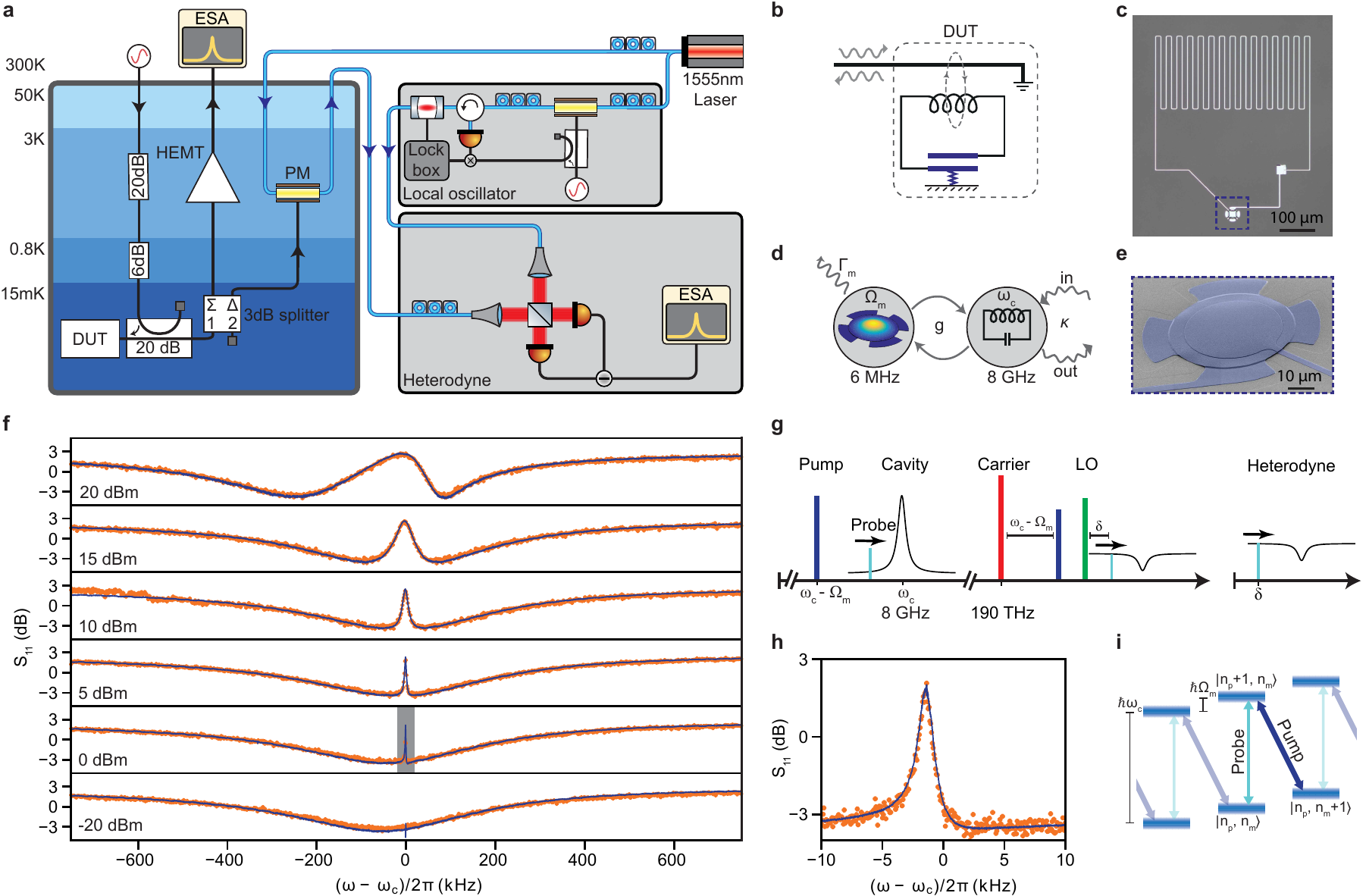}
	\caption{
		\footnotesize \linespread{1}
		\textbf{Electro-optic readout of a coherent microwave spectrum of a superconducting electromechanical system}.
		\textbf{a}, Experimental setup. Left: dilution fridge, right: optical setup.
		\textbf{b}, Electromechanical system used as a DUT.
		\textbf{c}, optical micrograph of the $LC$ resonator.
		\textbf{d}, Modal diagram of the electromechanical system.
		\textbf{e}, scanning electron micrograph of the	 mechanically compliant capacitor.
		\textbf{f}, coherent measurement of the electromechanical resonance for increasing microwave pump powers of $(-20, 0, 5, 10, 15, 20)\unit{dBm}$ at the source, from bottom to top respectively. The probe power is $-20~\unit{dBm}$ at the source.
		By increasing the pump power, the optomechanically induced transparency window emerges, and
		at stronger pump powers the modes get strongly coupled, leading to an avoided crossing effect.
		Blue lines correspond to HEMT readout and orange dots to optical readout.
		\textbf{g}, the frequency scheme for microwave tones, optical tones, and measured signal after heterodyning.
		\textbf{h}, high resolution measurement of the transparency window highlighted in~\textbf{f} with the gray box.
		\textbf{i}, level scheme of the optomechanical system. The pump tone is tuned close to red sideband transitions, in which a mechanical excitation quantum is annihilated (mechanical occupation
		$n_m\rightarrow n_m-1$) when a photon is added to the cavity (optical occupation $n_p\rightarrow n_p+1$), therefore coupling the corresponding energy eigenstates.
		The probe tone probes reflection in which the mechanical oscillator occupation is unchanged.
		The pump tone modifies the response of the cavity and creates a transparency window appears on resonance (OMIT).}
	\label{fig:3}
\end{figure*}

To further assess the performance of the PM at $800\unit{mK}$, we also performed a basic 
telecommunication experiment shown in Fig.~\ref{fig:2}e.
An arbitrary waveform generator (AWG) directly drives the PM with a pseudo-random bit sequence at a rate of $5\unit{GBaud}$.
We beat the optical phase-modulated carrier output with its reference arm, 
effectively forming a Mach-Zehnder interferometer whose average transmission is tuned to the quadrature point by adjusting the laser frequency,
and detect the electrical signal on the oscilloscope.
Figure~\ref{fig:2}f shows an eye diagram obtained from $8\times 10^5$ samples.
The open eye diagram features no error bits, hence the upper bound on bit error ratio is limited by total amount of measured samples and can be estimated to be
$5\times 10^{-5}$ with $95\%$ confidence level~\cite{BER2018}.
These measurements clearly demonstrate that the cryogenic modulator still functions at $800\unit{mK}$.

\subsection*{Optical Readout of Coherent Microwave Spectroscopy}
Having established the cryogenic modulation properties, we next carry out a cryogenic interconnect experiment, where the microwave output of a DUT is read out optically.
As an example system we employ a superconducting electro\-mechanical device in the form of a mechanically-compliant vacuum gap capacitor parametrically coupled to a superconducting microwave resonator (Fig.~\ref{fig:3}a--e).
These devices have been employed in a range of quantum electromechanical experiments, such as cooling the mechanical resonator to its quantum ground state~\cite{Teufel2011}, strong coupling between mechanical and microwave modes~\cite{TeufelStrongCoupling}, squeezing of mechanical motion~\cite{Wollman2015}, and demonstration of the quantum entanglement in the mechanical motion~\cite{ockeloen2018stabilized, barzanjeh2019stationary}, as well as implementing mechanically mediated tunable microwave non-reciprocity~\cite{Bernier2017} and quantum reservoir engineering~\cite{Toth2017}.
The microwave resonance (frequency $\omega_c\simeq 2\pi\times 8.2\unit{GHz}$ and linewidth $\kappa\simeq 2\pi\times 3\unit{MHz}$) is coupled to the mechanical resonance (frequency $\Omega_m\simeq 2\pi\times 6\unit{MHz}$ and linewidth $\Gamma_m\simeq 2\pi\times 10\unit{Hz}$) of the capacitor via electro\-mechanical coupling~\cite{RMP_optomechanics} (Fig.~\ref{fig:3}d).
The electro\-mechanical coupling rate is $g=g_0\sqrt{\bar{n}\s{cav}}$, where $g_0\simeq 2\pi\times150\unit{Hz}$ is independently characterized~\cite{NathanPhD} and $\bar{n}\s{cav}$ is intracavity microwave photon number, proportional to the microwave pump power. 
The system is inductively coupled to a microwave feed-line, enabling us to pump and read out the microwave mode in reflection. 

To demonstrate the electro-optical readout technique, we perform two-tone spectroscopy and measure optomechanically induced transparency (OMIT)~\cite{Weis2010,Zhou2013,safavi2011electromagnetically} (Fig.~\ref{fig:3}i) on the electromechanical sample, by applying a microwave pump tone on the lower motional sideband (red-detuned by $\Omega_m$ from the cavity resonance) and sweeping a second probe tone across the resonance.
The strong pump damps the mechanical motion, resulting in a wider effective mechanical linewidth, $\Gamma\s{eff}=\Gamma_m+4g^2/\kappa$.
The microwave pump modifies the cavity response due to the electromechanical coupling, resulting in a transparency window of width $\Gamma\s{eff}$ that appears on resonance, which we observe by the probe (Fig.~\ref{fig:3}g).
We performed an OMIT experiment for different pump powers and observed the mechanical resonance via the transparency feature.
In order to electro-optically read out the coherent response, the optical output is detected in a balanced heterodyne detector, using a frequency-shifted local oscillator (Fig.~\ref{fig:3}g).
Note that this scheme allows resolving spectroscopic features finer than the laser linewidth (Fig.~\ref{fig:3}h).
To compare the optical and HEMT readouts, the reflected signal is split and measured simultaneously using both techniques (Fig.~\ref{fig:3}a).
Figure~\ref{fig:3}f shows the OMIT results, with excellent agreement between the optical and HEMT readouts.
At high pump powers, when $g\sim\kappa$, we observe mode splitting as a result of strong coupling and mode-hybridization between the mechanical and microwave modes~\cite{TeufelStrongCoupling}.

\begin{figure}
	\includegraphics{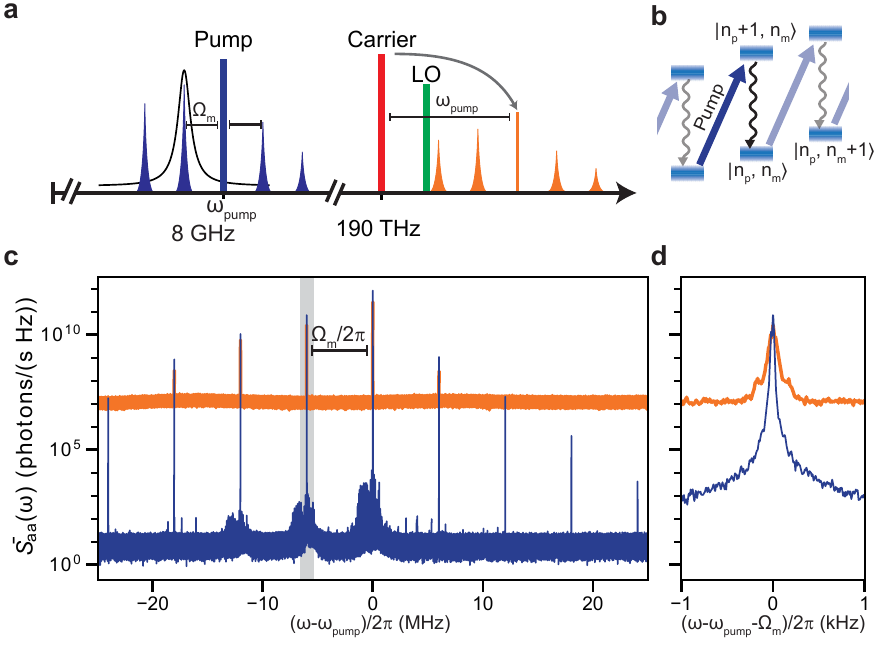}
	\caption{\footnotesize \linespread{1}
		\textbf{Electro-optic readout of an incoherent microwave spectrum of a superconducting electromechanical system}.
		\textbf{a}, Frequency-domain picture: a microwave tone pumps the electromechanical system on the upper motional sideband, inducing a parametric instability and generating mechanical sidebands equally spaced around the tone by the mechanical resonance frequency $\Omega_m$.
		The phase modulator transfers the microwave spectrum on the optical signal, which is subsequently mixed with a local oscillator (LO) and detected via heterodyne detection.
		\textbf{b}, Level scheme showing electromechanically induced parametric instability.
		A blue-detuned pump photon scatters into an on resonance photon and generates a phonon in the mechanical oscillator and causes anti-damping.
		At pump power above a certain threshold induces instability in the mechanical oscillator.
		\textbf{c}, Measured power spectral densities of the microwave pump (central peak) and mechanical sidebands detected by the HEMT (blue) and optical (orange) readouts.
		\textbf{d}, Enlargement of the gray-shaded area in \textbf{c}, showing the power spectral densities of the on-resonance mechanical sideband. 
		\label{fig:4}
	}
\end{figure}

\subsection*{Optical Readout of an Incoherent Microwave Spectrum}
Next, we employ our scheme to directly read out optically the power spectral density of a microwave signal emitted by the DUT.
For this, we drive the mechanical oscillator into self-oscillation by pumping the system on its upper motional sideband, $\omega\s{pump}=\omega_c+\Omega_m$, inducing a parametric instability~\cite{RMP_optomechanics,marquardt2006dynamical,carmon2005temporal,cattiaux2020beyond}.
The output microwave spectrum features strong sidebands around the microwave pump, at integer multiples of the mechanical frequency (Fig.~\ref{fig:4}a,b).
%
%
Figure~\ref{fig:4}c shows these mechanical signals obtained simultaneously using both our optical readout and the HEMT amplifier.
We use the known properties of the HEMT to estimate the transduction gain $G$ [Eq.~\eqref{eq:input_output_flux}] of our optical readout.
The blue trace in Fig.~\ref{fig:4}c shows the HEMT output referred back to its input using its known added noise, $n\s{add}^\mathrm{HEMT}\simeq 8\unit{quanta/(s\cdot Hz)}$, characterized independently.
This calibration yields the HEMT input signal $S$, which is equal to the PM microwave input.
The noise in the optically detected spectrum, referred to the optical output of the PM, is dominated by the optical shot noise, $1\unit{quanta/(s\cdot Hz)}$, for our $G\ll 1$.
In this calibration, we can obtain $G$ from the optical spectrum containing the transduced microwave signal $GS$.
The orange trace in Fig.~\ref{fig:4}c shows the optical noise spectrum referred to the microwave input, and Fig.~\ref{fig:4}d shows a zoom-in of a single sideband.
In this calibration, the signal areas in both measurements are equal to $S$.
Further explanation of this calibration is given in the appendix.
%
This yields 
$G=0.9\times 10^{-7}$, 
in good agreement with the theoretical value $G^\mathrm{theory}=3.5\times10^{-7}$ obtained from Eq.~\eqref{eq:G} using the measured output optical power, $P\s{opt}=1.1\unit{mW}$ (optical efficiency of $5\%$, including losses in fiber connectors and heterodyne detection setup).
We note that the frequency widening of the optically detected sidebands, observed in Fig.~\ref{fig:4}d, is due to fluctuations in the LO frequency, caused by the limited bandwidth of the locking setup in conjunction with using a minimal resolution bandwidth (RBW) of $1\unit{Hz}$ in the spectrum measurement.
The integrated sideband power, however, is conserved.
Improving the LO locking setup can reduce this effect. 

The added noise in the transduction process is (see Appendix)
\begin{equation}
n\s{add} = \frac{1}{2G} + n\s{th}^\mathrm{MW}+\frac{1}{2},
\label{eq:noise}
\end{equation}
where $n\s{th}^\mathrm{MW}$ is the average occupation of the thermal photonic bath due to the microwave fields.
This gives $n\s{add} \approx 6\times 10^{6}$.
The noise floor of the optical measurement in Fig.~\ref{fig:4}c is $60\unit{dB}$ above the HEMT readout.
This is due to the very small gain $G\sim 10^{-7}$, caused by the large $\Vpi$ and the limited optical power.
However, there is much room for improvement in these parameters.
Ref.~\onlinecite{Abel2019} reported a $\Vpi$-length product of $0.45\unit{V\,cm}$ in a BaTiO$_3$-based modulator, thus $\Vpi\sim 50\unit{mV}$ can be realized in a $\sim 10\unit{cm}$ device, possibly using low-loss superconducting electrodes~\cite{Yoshida1999,holzgrafe2020}.
The optical power can be increased arbitrarily in principle, however one needs to consider optical losses (mainly at the fiber-to-chip interfaces) that lead to heating.
Considering a device with an improved optical transmission of 66\%~\footnote{Commercial PMs with typical insertion loss $<2\unit{dB}$, ($>63\%$ transmission), at room temperature are already available.}
and incident power of $15\unit{mW}$, yields $P\s{opt}\sim 10\unit{mW}$.
This scenario achieves $G\sim 5\times 10^{-2}$ [Eq.~\eqref{eq:G}], with $n\s{add}\approx 20$ at $3\unit{K}$, competitive with HEMT performance, while the heat load of $5\unit{mW}$ is half that of a typical cryogenic HEMT.

%
%

It is worth mentioning that
many experiments utilize a near-quantum-limited pre-amplifier at the $15\unit{mK}$ stage (Fig.~\ref{fig:1}a,b).
In this case, the noise added in the second amplification stage, referred to the input, is $\sim\!(G\s{PA}G)^{-1}$ (see Appendix), where $G\s{PA}\sim 10^3$ is the pre-amplifier gain~\cite{yamamoto2008flux, macklin2015near, siddiqi2004rf, castellanos2008amplification}.
Thus, $G\gtrsim G\s{PA}^{-1}$ suffices to preserve near-quantum-limited amplification (See Appendix).

\subsection*{Conclusions}
We have demonstrated the viability of LiNbO$_3$ devices, currently-employed in the telecommunication market, as electro-optical interconnects in cryogenic platforms used in superconducting quantum technologies, in particular as viable alternative to HEMT amplifiers with the potential of reduced heat load.
By interfacing a commercial PM to a circuit-electromechanical system 
that was previously used to perform quantum experiments, we implemented an electro-optical readout of this system. In addition, we quantified the gap between conventional microwave amplifiers and the electro-optical alternative.
It is feasible that this gap be closed in the near future, by improved devices with lower $\Vpi$, resulting in a near-quantum-limited broadband microwave-to-optical interconnect.

\appendix*
\section{Quantum mechanical model for a phase modulator}
In the following, 
we derive a simple quantum description of the phase modulator
to establish the quantum limits in transducing the input microwaves.
The central assumption is that the linear regime stays valid,
for sufficiently low input microwave powers.
As such, the scattering equations linking inputs to output
should be identical in both quantum and classical cases.
We can use the known classical regime as a starting point, with
the output optical field amplitude 
$\hat{a}_\out$
expressed
as a function of the input optical field
$\hat{a}_\inp$
as
\begin{equation}
\hat{a}_\out = e^{- i\pi V/\Vpi}
\hat{a}_\inp \approx (1 - i \pi V/\Vpi)\hat{a}_\inp,
\label{eq:cl_lin}
\end{equation}
where
$V$ is the classical voltage applied at the input and
the half-wave voltage $\Vpi$ is the voltage at which the phase modulator applies a phase shift of 
$\pi$.
For the quantum model, the classical fields are replaced by their quantum equivalent.
The microwave input becomes
$\hat V = 
\sqrt{\hbar \omega\s{MW} Z_0}
(\hat{b} + \hat{b}^\dagger)/{\sqrt{2}}$
with $\hat{b}$ the annihilation operator for the microwave field at frequency $\omega\s{MW}$ traveling on a transmission line of impedance $Z_0$.
The optical input is $\hat{a}_\inp = \alpha e^{-i\omega\s{opt} t} + \dain$, where
$\alpha$ is the amplitude of the coherent carrier field of frequency $\omega\s{opt}$,
with $\lvert\alpha\rvert^2 = P\s{opt}/\hbar\omega\s{opt}$,
and $\dain$ carries the quantum fluctuations of the input optical field.
Inserting the expressions in Eq.~\eqref{eq:cl_lin}, we can compute 
$\daout = \hat{a}_\out - \alpha e^{-i \omega_\mathrm{opt} t}$, the quantum fluctuations of the output optical field,
given by
\begin{equation}
\daout
=
\delta a_\inp - i \sqrt G e^{-i \omega_\mathrm{opt} t} (\hat{b} + \hat{b}^\dagger)
\label{eq:qu_opt}
\end{equation}
with the transduction gain $G$ given by Eq.~\eqref{eq:G}.

To understand the implications of Eq.~\eqref{eq:qu_opt} for the quantum noise in the transduction,
we compute the power spectral density of the output optical field,
\begin{equation}
\begin{aligned}
\sd_{\modedadag \modeda}^\out [\omega\s{opt} + \omega\s{MW}]=& \sd_{\delta a^\dagger \delta a}^\inp [\omega\s{opt} + \omega\s{MW}]\\
&+ G(\sd_{b^\dagger b}[\omega\s{MW}] + \sd_{b b^\dagger}[-\omega\s{MW}]).
\end{aligned}
\end{equation}
The first term corresponds to the \emph{added} quantum noise due to the 
input optical field. 
The second term contains contributions from the microwave frequency $\omega\s{MW}$, including both the signal and noise. 
The third term contains \emph{added} microwave noise at frequency $-\omega\s{MW}$, composed of thermal and quantum noise components, respectively $n^\mathrm{MW}\s{th}+1/2$.
%
%
Thus Eq.~\eqref{eq:sd_out} can be simplified to
\begin{equation}
\sd_{\modedadag \modeda}^\out [\omega\s{opt} + \omega\s{MW}] = 
G\,\sd_{b^\dagger b}[\omega\s{MW}]
+ G\biggl(n^\mathrm{MW}\s{th} + \frac{1}{2}\biggr)
+ \frac{1}{2}.
\label{eq:Full_conversion}
\end{equation}
We emphasize two limiting cases.
When $G\ll 1$, as in our experiment, the added noise is dominated by the input optical quantum noise, the last term in Eq.~\eqref{eq:Full_conversion}.
In the opposite limit, $G\gg 1$, the added noise is dominated by the microwave input noise, and the signal-to-noise ratio is independent of $G$.
In any case, the added noise referred to the input is given by Eq.~\eqref{eq:noise}.

\section{Calibration of the transduction gain.}
\begin{figure}
	\includegraphics[scale=1]{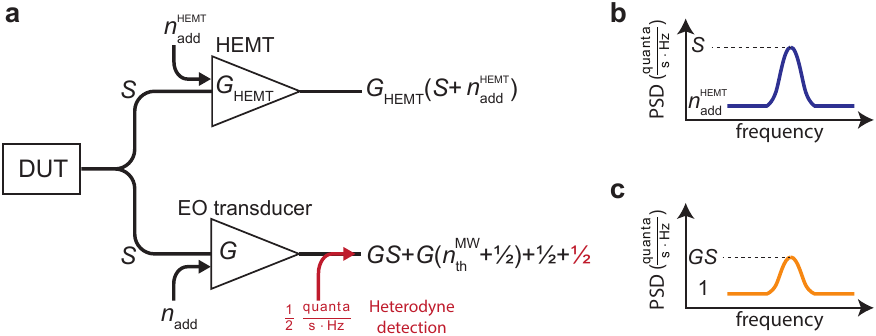}
	\caption{\footnotesize \linespread{1}
		\textbf{Illustration of the gain characterization procedure.}
		\textbf{a,} Propagation of the DUT signal through the system.
		\textbf{b,} Power spectral density of the HEMT output.
		\textbf{c,} Power spectral density of the optical heterodyne detector.
		\label{fig:gain_charac}}
\end{figure}

Figure~\ref{fig:gain_charac} illustrates the procedure of experimentally characterizing the transduction gain of our electro-optic transducer.
The microwave signal is split equally into two parts $S$, fed to the HEMT amplifier and the PM respectively (Fig.~\ref{fig:gain_charac}a).
The HEMT added noise is characterized independently to be $n^\mathrm{HEMT}\s{add} = 8\unit{quanta/(s\cdot Hz)}$.
We infer $S$ from the spectrum of the HEMT amplified signal, referring the noise floor to $n^\mathrm{HEMT}\s{add}$ (Fig.~\ref{fig:gain_charac}b).
In the PM branch, the added noise of the transduction is given by Eq.~\eqref{eq:noise}.
The spectrum is detected using a balanced heterodyne detector, which adds  $1/2\unit{quanta/(s\cdot Hz)}$ of noise (Fig.~\ref{fig:gain_charac}c).
We can safely neglect $G(n\s{MW}+1/2)$ and consider the noise floor of the spectrum referred to the input of the heterodyne detector, i.e. $1\unit{quanta/(s\cdot Hz)}$.
This allows us to calculate $GS$, and finally obtain $G$ with knowledge of $S$ from the HEMT measurement.

\begin{figure}
	\includegraphics[scale=1]{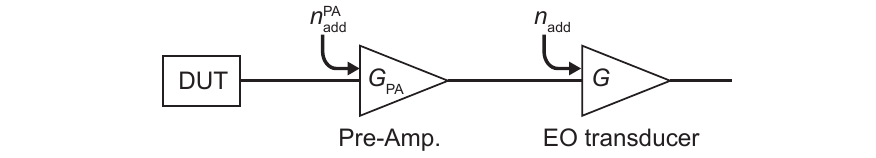}
	\caption{\footnotesize \linespread{1} Schematic signal flow when pre-amplifier is used.
		\label{fig:preamp}}
\end{figure}

When using a quantum-limited pre-amplifiers before the electro-optical transducer (not done in our experiment), we can model the readout chain as shown in Fig.~\ref{fig:preamp}b.
The total added noise of the readout chain is
\begin{equation}
n\s{add}^\mathrm{total} = n\s{add}^\mathrm{PA} + \frac{n\s{add}}{G\s{PA}} \simeq n\s{add}^\mathrm{PA} + \frac{1}{2 G\s{PA}G}
\end{equation}
Therefore when $G\simeq 1/G\s{PA}$, the total added noise will be dominated by $n\s{add}^\mathrm{PA}\sim 1\unit{quanta/(s\cdot Hz)}$~\cite{yamamoto2008flux, macklin2015near, siddiqi2004rf, castellanos2008amplification} and the readout will be near-quantum-limited.

\section{Experimental details and heating measurements.}
We use a fiber-coupled lithium niobate PM from Thorlabs, model LN65S, used as-is with no modifications.
Note that the minimum specified operating temperature is $0\unit{^\circ C}$.
The device sustained several cooldown-warmup cycles with reversible behavior in its optical transmission. 
We measured 25\% reduction in the optical transmission at cryogenic relative to room temperature.
The PM metallic box was tightly clamped to the flange of the $800\unit{mK}$ or $3\unit{K}$ stage.
We use a Bluefors LD-250 dilution refrigerator. The approxmiate \textit{available} cooling powers of the $\{ 15\unit{mK}, 800\unit{mK}, 3\unit{K}\}$ stages are $\{ 12\unit{\mu W}, 30\unit{mW}, 300\unit{mW}\}$.

\begin{figure*}[ht]
	\includegraphics[width=\textwidth]{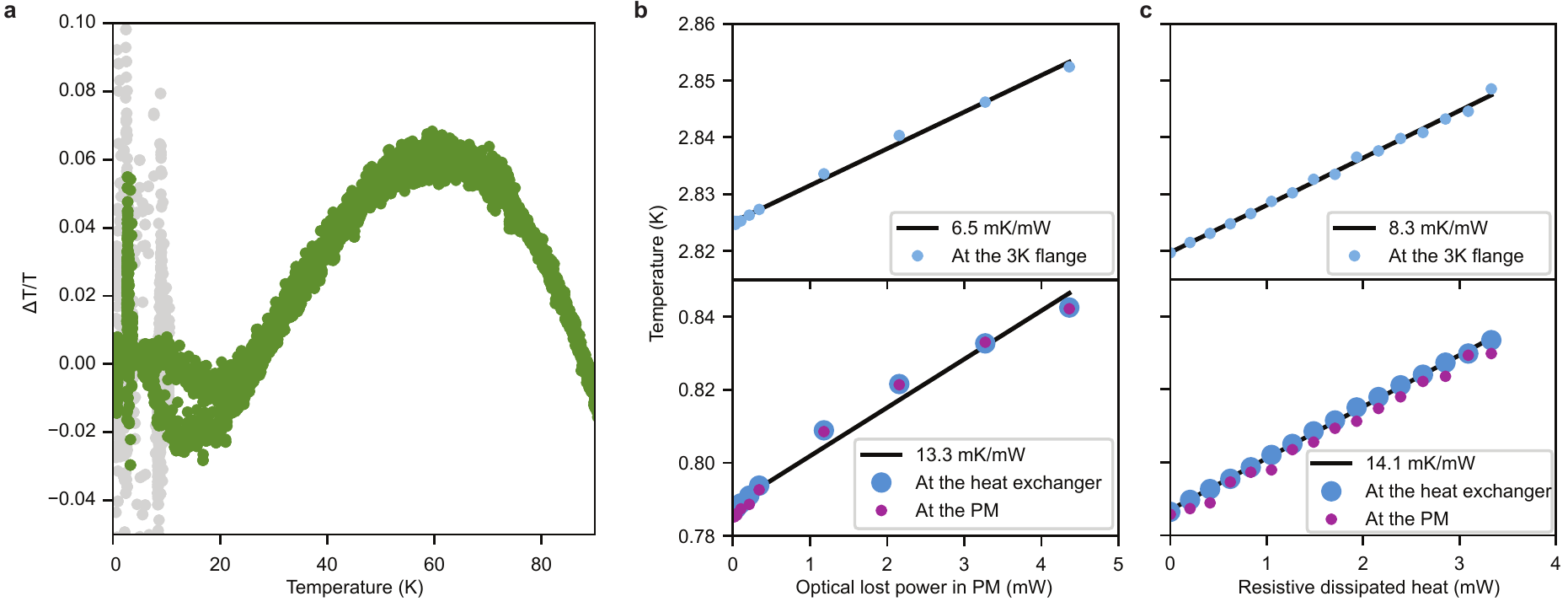}
	\caption{
		\footnotesize \linespread{1}
		\textbf{Heat dissipation and temperature gradients}.
		\textbf{a}, Relative temperature difference between PM box and heat exchanger, on $800\unit{mK}$ flange during a cooldown.
		The gray datapoints correpsond to specific periods of pulse precooling and mixture condensation, where the temperature is unstable.
		\textbf{b}, Measurement of heating due to optical dissipation when the phase modulator is mounted on the $800\unit{mK}$ flange.
		\textbf{c}, Measurement of heating using a calibrated resistive heater mounted on the $800\unit{mK}$ flange.
	}
	\label{fig:8_SI}
\end{figure*}

Figure~\ref{fig:2}c shows the variation of $\Vpi$ from room temperature to $800\unit{mK}$, obtained during a cooldown of the dilution fridge and measured using the default thermometer on the $800\unit{mK}$ flange, located next to the heat exchanger, about $10\unit{cm}$ from the PM.
In order to rule out possible temperature gradients, we mounted a calibrated thermometer next to the PM and monitored both thermometer readings during cooldown.
Figure~\ref{fig:8_SI}a shows the measured relative temperature difference, which is less than $\sim 5\%$ throughout the cooldown. Note that this excludes pulse precooling and mixture condensation period when the temperature is unstable (shown for completeness in Fig.~\ref{fig:8_SI}a).

Figure~\ref{fig:2}d shows the temperature increase of the $15\unit{mK}$, $800\unit{mK}$, and $3\unit{mK}$ stages of the dilution fridge as a function of the optical power incident on the PM, which is mounted on the $800\unit{mK}$ stage.
We performed a simple measurement 
to verify that this temperature increase can be ascribed to light absorbed in the PM body (and not, e.g, light leakage into the fridge volume), corresponding to the optical transmission (insertion loss) of the PM.
We used the calibrated $120\unit{\Omega}$ still heater built in the $800\unit{mK}$ stage to apply heat directly, we then repeated the measurement using optical input to the PM as the heating source (as in Fig.~\ref{fig:2}d).
Figure \ref{fig:8_SI}a,b compares the results of this measurement, showing temperature increase in the $3\unit{K}$ and $800\unit{mK}$ stages (the latter recorded with the two separated thermometers) vs.~dissipated power.
In the case of optical heating, the dissipated power is computed directly from the incident power on the PM and its measured transmission of $23\%$.

Figure~\ref{fig:8_SI}b,c shows the result of this measurement.
The optical heating shows a temperature increase of
$13.3\unit{mK/mW}$ ($6.5\unit{mK/mW}$) at the $800\unit{mK}$ ($3\unit{K}$) stage (Fig.\ref{fig:8_SI}a),
while the resistive heating shows a temperature increase of 
$14.1\unit{mK/mW}$ ($8.3\unit{mK/mW}$) at the $800\unit{mK}$ ($3\unit{K}$) stage (Fig.\ref{fig:8_SI}b).
Thus heating due to operation of the electro-optical interconnect is very similar to localized, resistive heating.

%

\subsection*{Data availability statement}
The code and data used to produce the plots within this paper will be available at a Zenodo open-access repository. All other data used in this study are available from the corresponding authors upon reasonable request.

\begin{acknowledgements}
We thank Nils J.~Engelsen for thorough reading of the manuscript.
This work was supported by the European Union's Horizon 2020 research and innovation programme under grant agreement No. 732894 (FET Proactive HOT), and from the European Research Council (ERC) under the European Union’s Horizon 2020 research and innovation programme (grant agreement No. 835329). This work was supported by funding from the Swiss National Science Foundation under grant agreement NCCR-QSIT: 51NF40\_185902 and Sinergia grant no. 186364 (QuantEOM). The circuit electro-mechanical device was fabricated in the Center of MicroNanoTechnology (CMi) at EPFL.
\end{acknowledgements}

\bibliography{sample}

\begin{thebibliography}{65}%
\makeatletter
\providecommand \@ifxundefined [1]{%
 \@ifx{#1\undefined}
}%
\providecommand \@ifnum [1]{%
 \ifnum #1\expandafter \@firstoftwo
 \else \expandafter \@secondoftwo
 \fi
}%
\providecommand \@ifx [1]{%
 \ifx #1\expandafter \@firstoftwo
 \else \expandafter \@secondoftwo
 \fi
}%
\providecommand \natexlab [1]{#1}%
\providecommand \enquote  [1]{``#1''}%
\providecommand \bibnamefont  [1]{#1}%
\providecommand \bibfnamefont [1]{#1}%
\providecommand \citenamefont [1]{#1}%
\providecommand \href@noop [0]{\@secondoftwo}%
\providecommand \href [0]{\begingroup \@sanitize@url \@href}%
\providecommand \@href[1]{\@@startlink{#1}\@@href}%
\providecommand \@@href[1]{\endgroup#1\@@endlink}%
\providecommand \@sanitize@url [0]{\catcode `\\12\catcode `\$12\catcode
  `\&12\catcode `\#12\catcode `\^12\catcode `\_12\catcode `\%12\relax}%
\providecommand \@@startlink[1]{}%
\providecommand \@@endlink[0]{}%
\providecommand \url  [0]{\begingroup\@sanitize@url \@url }%
\providecommand \@url [1]{\endgroup\@href {#1}{\urlprefix }}%
\providecommand \urlprefix  [0]{URL }%
\providecommand \Eprint [0]{\href }%
\providecommand \doibase [0]{http://dx.doi.org/}%
\providecommand \selectlanguage [0]{\@gobble}%
\providecommand \bibinfo  [0]{\@secondoftwo}%
\providecommand \bibfield  [0]{\@secondoftwo}%
\providecommand \translation [1]{[#1]}%
\providecommand \BibitemOpen [0]{}%
\providecommand \bibitemStop [0]{}%
\providecommand \bibitemNoStop [0]{.\EOS\space}%
\providecommand \EOS [0]{\spacefactor3000\relax}%
\providecommand \BibitemShut  [1]{\csname bibitem#1\endcsname}%
\let\auto@bib@innerbib\@empty
\bibitem [{\citenamefont {Winzer}\ \emph {et~al.}(2018)\citenamefont {Winzer},
  \citenamefont {Neilson},\ and\ \citenamefont {Chraplyvy}}]{winzer2018fiber}%
  \BibitemOpen
  \bibfield  {author} {\bibinfo {author} {\bibfnamefont {Peter~J}\ \bibnamefont
  {Winzer}}, \bibinfo {author} {\bibfnamefont {David~T}\ \bibnamefont
  {Neilson}}, \ and\ \bibinfo {author} {\bibfnamefont {Andrew~R}\ \bibnamefont
  {Chraplyvy}},\ }\bibfield  {title} {\enquote {\bibinfo {title} {Fiber-optic
  transmission and networking: the previous 20 and the next 20 years},}\ }\href
  {https://www.osapublishing.org/oe/abstract.cfm?uri=oe-26-18-24190} {\bibfield
   {journal} {\bibinfo  {journal} {Optics express}\ }\textbf {\bibinfo {volume}
  {26}},\ \bibinfo {pages} {24190--24239} (\bibinfo {year} {2018})}\BibitemShut
  {NoStop}%
\bibitem [{\citenamefont {Kachris}\ and\ \citenamefont
  {Tomkos}(2012)}]{kachris2012survey}%
  \BibitemOpen
  \bibfield  {author} {\bibinfo {author} {\bibfnamefont {Christoforos}\
  \bibnamefont {Kachris}}\ and\ \bibinfo {author} {\bibfnamefont {Ioannis}\
  \bibnamefont {Tomkos}},\ }\bibfield  {title} {\enquote {\bibinfo {title} {A
  {Survey} on {Optical} {Interconnects} for {Data} {Centers}},}\ }\href
  {\doibase 10.1109/SURV.2011.122111.00069} {\bibfield  {journal} {\bibinfo
  {journal} {IEEE Communications Surveys Tutorials}\ }\textbf {\bibinfo
  {volume} {14}},\ \bibinfo {pages} {1021--1036} (\bibinfo {year}
  {2012})}\BibitemShut {NoStop}%
\bibitem [{\citenamefont {Devoret}\ and\ \citenamefont
  {Schoelkopf}(2013)}]{devoret_superconducting_2013}%
  \BibitemOpen
  \bibfield  {author} {\bibinfo {author} {\bibfnamefont {M.~H.}\ \bibnamefont
  {Devoret}}\ and\ \bibinfo {author} {\bibfnamefont {R.~J.}\ \bibnamefont
  {Schoelkopf}},\ }\bibfield  {title} {\enquote {\bibinfo {title}
  {Superconducting {Circuits} for {Quantum} {Information}: {An} {Outlook}},}\
  }\href {\doibase 10.1126/science.1231930} {\bibfield  {journal} {\bibinfo
  {journal} {Science}\ }\textbf {\bibinfo {volume} {339}},\ \bibinfo {pages}
  {1169--1174} (\bibinfo {year} {2013})}\BibitemShut {NoStop}%
\bibitem [{\citenamefont {Martinis}\ \emph {et~al.}(2020)\citenamefont
  {Martinis}, \citenamefont {Devoret},\ and\ \citenamefont
  {Clarke}}]{martinis_quantum_2020}%
  \BibitemOpen
  \bibfield  {author} {\bibinfo {author} {\bibfnamefont {John~M.}\ \bibnamefont
  {Martinis}}, \bibinfo {author} {\bibfnamefont {Michel~H.}\ \bibnamefont
  {Devoret}}, \ and\ \bibinfo {author} {\bibfnamefont {John}\ \bibnamefont
  {Clarke}},\ }\bibfield  {title} {\enquote {\bibinfo {title} {Quantum
  {Josephson} junction circuits and the dawn of artificial atoms},}\ }\href
  {\doibase 10.1038/s41567-020-0829-5} {\bibfield  {journal} {\bibinfo
  {journal} {Nature Physics}\ }\textbf {\bibinfo {volume} {16}},\ \bibinfo
  {pages} {234--237} (\bibinfo {year} {2020})}\BibitemShut {NoStop}%
\bibitem [{\citenamefont {Blais}\ \emph {et~al.}(2020)\citenamefont {Blais},
  \citenamefont {Girvin},\ and\ \citenamefont {Oliver}}]{blais_quantum_2020}%
  \BibitemOpen
  \bibfield  {author} {\bibinfo {author} {\bibfnamefont {Alexandre}\
  \bibnamefont {Blais}}, \bibinfo {author} {\bibfnamefont {Steven~M.}\
  \bibnamefont {Girvin}}, \ and\ \bibinfo {author} {\bibfnamefont {William~D.}\
  \bibnamefont {Oliver}},\ }\bibfield  {title} {\enquote {\bibinfo {title}
  {Quantum information processing and quantum optics with circuit quantum
  electrodynamics},}\ }\href {\doibase 10.1038/s41567-020-0806-z} {\bibfield
  {journal} {\bibinfo  {journal} {Nature Physics}\ }\textbf {\bibinfo {volume}
  {16}},\ \bibinfo {pages} {247--256} (\bibinfo {year} {2020})}\BibitemShut
  {NoStop}%
\bibitem [{\citenamefont {Clerk}\ \emph {et~al.}(2020)\citenamefont {Clerk},
  \citenamefont {Lehnert}, \citenamefont {Bertet}, \citenamefont {Petta},\ and\
  \citenamefont {Nakamura}}]{clerk_hybrid_2020}%
  \BibitemOpen
  \bibfield  {author} {\bibinfo {author} {\bibfnamefont {A.~A.}\ \bibnamefont
  {Clerk}}, \bibinfo {author} {\bibfnamefont {K.~W.}\ \bibnamefont {Lehnert}},
  \bibinfo {author} {\bibfnamefont {P.}~\bibnamefont {Bertet}}, \bibinfo
  {author} {\bibfnamefont {J.~R.}\ \bibnamefont {Petta}}, \ and\ \bibinfo
  {author} {\bibfnamefont {Y.}~\bibnamefont {Nakamura}},\ }\bibfield  {title}
  {\enquote {\bibinfo {title} {Hybrid quantum systems with circuit quantum
  electrodynamics},}\ }\href {\doibase 10.1038/s41567-020-0797-9} {\bibfield
  {journal} {\bibinfo  {journal} {Nature Physics}\ }\textbf {\bibinfo {volume}
  {16}},\ \bibinfo {pages} {257--267} (\bibinfo {year} {2020})}\BibitemShut
  {NoStop}%
\bibitem [{\citenamefont {Wooten}\ \emph {et~al.}(2000)\citenamefont {Wooten},
  \citenamefont {Kissa}, \citenamefont {Yi-Yan}, \citenamefont {Murphy},
  \citenamefont {Lafaw} \emph {et~al.}}]{wooten2000review}%
  \BibitemOpen
  \bibfield  {author} {\bibinfo {author} {\bibfnamefont {Ed~L}\ \bibnamefont
  {Wooten}}, \bibinfo {author} {\bibfnamefont {Karl~M}\ \bibnamefont {Kissa}},
  \bibinfo {author} {\bibfnamefont {Alfredo}\ \bibnamefont {Yi-Yan}}, \bibinfo
  {author} {\bibfnamefont {Edmond~J}\ \bibnamefont {Murphy}}, \bibinfo {author}
  {\bibfnamefont {Donald~A}\ \bibnamefont {Lafaw}},  \emph {et~al.},\
  }\bibfield  {title} {\enquote {\bibinfo {title} {A review of lithium niobate
  modulators for fiber-optic communications systems},}\ }\href
  {https://ieeexplore.ieee.org/abstract/document/826874} {\bibfield  {journal}
  {\bibinfo  {journal} {IEEE Journal of selected topics in Quantum
  Electronics}\ }\textbf {\bibinfo {volume} {6}},\ \bibinfo {pages} {69--82}
  (\bibinfo {year} {2000})}\BibitemShut {NoStop}%
\bibitem [{\citenamefont {Pospieszalski}\ \emph {et~al.}(1988)\citenamefont
  {Pospieszalski}, \citenamefont {Weinreb}, \citenamefont {Norrod},\ and\
  \citenamefont {Harris}}]{pospieszalski1988fets}%
  \BibitemOpen
  \bibfield  {author} {\bibinfo {author} {\bibfnamefont {Marian~W}\
  \bibnamefont {Pospieszalski}}, \bibinfo {author} {\bibfnamefont {Sander}\
  \bibnamefont {Weinreb}}, \bibinfo {author} {\bibfnamefont {Roger~D}\
  \bibnamefont {Norrod}}, \ and\ \bibinfo {author} {\bibfnamefont {Ronald}\
  \bibnamefont {Harris}},\ }\bibfield  {title} {\enquote {\bibinfo {title}
  {Fets and hemts at cryogenic temperatures-their properties and use in
  low-noise amplifiers},}\ }\href
  {https://ieeexplore.ieee.org/abstract/document/3548} {\bibfield  {journal}
  {\bibinfo  {journal} {IEEE transactions on microwave theory and techniques}\
  }\textbf {\bibinfo {volume} {36}},\ \bibinfo {pages} {552--560} (\bibinfo
  {year} {1988})}\BibitemShut {NoStop}%
\bibitem [{\citenamefont {Duh}\ \emph {et~al.}(1988)\citenamefont {Duh},
  \citenamefont {Pospieszalski}, \citenamefont {Kopp}, \citenamefont {Ho},
  \citenamefont {Jabra} \emph {et~al.}}]{duh1988ultra}%
  \BibitemOpen
  \bibfield  {author} {\bibinfo {author} {\bibfnamefont {KH~George}\
  \bibnamefont {Duh}}, \bibinfo {author} {\bibfnamefont {Marian~W}\
  \bibnamefont {Pospieszalski}}, \bibinfo {author} {\bibfnamefont {William~F}\
  \bibnamefont {Kopp}}, \bibinfo {author} {\bibfnamefont {Pin}\ \bibnamefont
  {Ho}}, \bibinfo {author} {\bibfnamefont {Amani~A}\ \bibnamefont {Jabra}},
  \emph {et~al.},\ }\bibfield  {title} {\enquote {\bibinfo {title}
  {Ultra-low-noise cryogenic high-electron-mobility transistors},}\ }\href
  {https://ieeexplore.ieee.org/abstract/document/2448} {\bibfield  {journal}
  {\bibinfo  {journal} {IEEE transactions on electron devices}\ }\textbf
  {\bibinfo {volume} {35}},\ \bibinfo {pages} {249--256} (\bibinfo {year}
  {1988})}\BibitemShut {NoStop}%
\bibitem [{\citenamefont {Weis}\ \emph {et~al.}(2010)\citenamefont {Weis},
  \citenamefont {Riviere}, \citenamefont {Deleglise}, \citenamefont {Gavartin},
  \citenamefont {Arcizet} \emph {et~al.}}]{Weis2010}%
  \BibitemOpen
  \bibfield  {author} {\bibinfo {author} {\bibfnamefont {S.}~\bibnamefont
  {Weis}}, \bibinfo {author} {\bibfnamefont {R.}~\bibnamefont {Riviere}},
  \bibinfo {author} {\bibfnamefont {S.}~\bibnamefont {Deleglise}}, \bibinfo
  {author} {\bibfnamefont {E.}~\bibnamefont {Gavartin}}, \bibinfo {author}
  {\bibfnamefont {O.}~\bibnamefont {Arcizet}},  \emph {et~al.},\ }\bibfield
  {title} {\enquote {\bibinfo {title} {Optomechanically induced
  transparency},}\ }\href {\doibase 10.1126/science.1195596} {\bibfield
  {journal} {\bibinfo  {journal} {Science}\ }\textbf {\bibinfo {volume}
  {330}},\ \bibinfo {pages} {1520–1523} (\bibinfo {year} {2010})}\BibitemShut
  {NoStop}%
\bibitem [{\citenamefont {Zhou}\ \emph {et~al.}(2013)\citenamefont {Zhou},
  \citenamefont {Hocke}, \citenamefont {Schliesser}, \citenamefont {Marx},
  \citenamefont {Huebl} \emph {et~al.}}]{Zhou2013}%
  \BibitemOpen
  \bibfield  {author} {\bibinfo {author} {\bibfnamefont {X.}~\bibnamefont
  {Zhou}}, \bibinfo {author} {\bibfnamefont {F.}~\bibnamefont {Hocke}},
  \bibinfo {author} {\bibfnamefont {A.}~\bibnamefont {Schliesser}}, \bibinfo
  {author} {\bibfnamefont {A.}~\bibnamefont {Marx}}, \bibinfo {author}
  {\bibfnamefont {H.}~\bibnamefont {Huebl}},  \emph {et~al.},\ }\bibfield
  {title} {\enquote {\bibinfo {title} {Slowing, advancing and switching of
  microwave signals using circuit nanoelectromechanics},}\ }\href {\doibase
  10.1038/nphys2527} {\bibfield  {journal} {\bibinfo  {journal} {Nature
  Physics}\ }\textbf {\bibinfo {volume} {9}},\ \bibinfo {pages} {179–184}
  (\bibinfo {year} {2013})}\BibitemShut {NoStop}%
\bibitem [{\citenamefont {Safavi-Naeini}\ \emph {et~al.}(2011)\citenamefont
  {Safavi-Naeini}, \citenamefont {Alegre}, \citenamefont {Chan}, \citenamefont
  {Eichenfield}, \citenamefont {Winger} \emph
  {et~al.}}]{safavi2011electromagnetically}%
  \BibitemOpen
  \bibfield  {author} {\bibinfo {author} {\bibfnamefont {Amir~H}\ \bibnamefont
  {Safavi-Naeini}}, \bibinfo {author} {\bibfnamefont {TP~Mayer}\ \bibnamefont
  {Alegre}}, \bibinfo {author} {\bibfnamefont {Jasper}\ \bibnamefont {Chan}},
  \bibinfo {author} {\bibfnamefont {Matt}\ \bibnamefont {Eichenfield}},
  \bibinfo {author} {\bibfnamefont {Martin}\ \bibnamefont {Winger}},  \emph
  {et~al.},\ }\bibfield  {title} {\enquote {\bibinfo {title}
  {Electromagnetically induced transparency and slow light with
  optomechanics},}\ }\href {https://www.nature.com/articles/nature09933}
  {\bibfield  {journal} {\bibinfo  {journal} {Nature}\ }\textbf {\bibinfo
  {volume} {472}},\ \bibinfo {pages} {69--73} (\bibinfo {year}
  {2011})}\BibitemShut {NoStop}%
\bibitem [{\citenamefont {Teufel}\ \emph
  {et~al.}(2011{\natexlab{a}})\citenamefont {Teufel}, \citenamefont {Li},
  \citenamefont {Allman}, \citenamefont {Cicak}, \citenamefont {Sirois} \emph
  {et~al.}}]{TeufelStrongCoupling}%
  \BibitemOpen
  \bibfield  {author} {\bibinfo {author} {\bibfnamefont {J.~D.}\ \bibnamefont
  {Teufel}}, \bibinfo {author} {\bibfnamefont {Dale}\ \bibnamefont {Li}},
  \bibinfo {author} {\bibfnamefont {M.~S.}\ \bibnamefont {Allman}}, \bibinfo
  {author} {\bibfnamefont {K.}~\bibnamefont {Cicak}}, \bibinfo {author}
  {\bibfnamefont {A.~J.}\ \bibnamefont {Sirois}},  \emph {et~al.},\ }\bibfield
  {title} {\enquote {\bibinfo {title} {Circuit cavity electromechanics in the
  strong-coupling regime},}\ }\href {\doibase 10.1038/nature09898} {\bibfield
  {journal} {\bibinfo  {journal} {Nature}\ }\textbf {\bibinfo {volume} {471}},\
  \bibinfo {pages} {204–208} (\bibinfo {year}
  {2011}{\natexlab{a}})}\BibitemShut {NoStop}%
\bibitem [{\citenamefont {Wang}\ \emph {et~al.}(2018)\citenamefont {Wang},
  \citenamefont {Zhang}, \citenamefont {Chen}, \citenamefont {Bertrand},
  \citenamefont {Shams-Ansari} \emph {et~al.}}]{Loncar2018}%
  \BibitemOpen
  \bibfield  {author} {\bibinfo {author} {\bibfnamefont {Cheng}\ \bibnamefont
  {Wang}}, \bibinfo {author} {\bibfnamefont {Mian}\ \bibnamefont {Zhang}},
  \bibinfo {author} {\bibfnamefont {Xi}~\bibnamefont {Chen}}, \bibinfo {author}
  {\bibfnamefont {Maxime}\ \bibnamefont {Bertrand}}, \bibinfo {author}
  {\bibfnamefont {Amirhassan}\ \bibnamefont {Shams-Ansari}},  \emph {et~al.},\
  }\bibfield  {title} {\enquote {\bibinfo {title} {Integrated lithium niobate
  electro-optic modulators operating at cmos-compatible voltages},}\ }\href
  {\doibase 10.1038/s41586-018-0551-y} {\bibfield  {journal} {\bibinfo
  {journal} {Nature}\ }\textbf {\bibinfo {volume} {562}},\ \bibinfo {pages}
  {101–104} (\bibinfo {year} {2018})}\BibitemShut {NoStop}%
\bibitem [{\citenamefont {He}\ \emph {et~al.}(2019)\citenamefont {He},
  \citenamefont {Yang}, \citenamefont {Ling}, \citenamefont {Luo},
  \citenamefont {Liang}, \citenamefont {Li}, \citenamefont {Shen},
  \citenamefont {Wang}, \citenamefont {Vahala},\ and\ \citenamefont
  {Lin}}]{he2019}%
  \BibitemOpen
  \bibfield  {author} {\bibinfo {author} {\bibfnamefont {Yang}\ \bibnamefont
  {He}}, \bibinfo {author} {\bibfnamefont {Qi-Fan}\ \bibnamefont {Yang}},
  \bibinfo {author} {\bibfnamefont {Jingwei}\ \bibnamefont {Ling}}, \bibinfo
  {author} {\bibfnamefont {Rui}\ \bibnamefont {Luo}}, \bibinfo {author}
  {\bibfnamefont {Hanxiao}\ \bibnamefont {Liang}}, \bibinfo {author}
  {\bibfnamefont {Mingxiao}\ \bibnamefont {Li}}, \bibinfo {author}
  {\bibfnamefont {Boqiang}\ \bibnamefont {Shen}}, \bibinfo {author}
  {\bibfnamefont {Heming}\ \bibnamefont {Wang}}, \bibinfo {author}
  {\bibfnamefont {Kerry}\ \bibnamefont {Vahala}}, \ and\ \bibinfo {author}
  {\bibfnamefont {Qiang}\ \bibnamefont {Lin}},\ }\bibfield  {title} {\enquote
  {\bibinfo {title} {Self-starting bi-chromatic {LiNbO}$_3$ soliton
  microcomb},}\ }\href {\doibase 10.1364/OPTICA.6.001138} {\bibfield  {journal}
  {\bibinfo  {journal} {Optica}\ }\textbf {\bibinfo {volume} {6}},\ \bibinfo
  {pages} {1138--1144} (\bibinfo {year} {2019})}\BibitemShut {NoStop}%
\bibitem [{\citenamefont {Thiele}\ \emph {et~al.}(2020)\citenamefont {Thiele},
  \citenamefont {Bruch}, \citenamefont {Quiring}, \citenamefont {Ricken},
  \citenamefont {Herrmann}, \citenamefont {Eigner}, \citenamefont
  {Silberhorn},\ and\ \citenamefont {Bartley}}]{thiele2020}%
  \BibitemOpen
  \bibfield  {author} {\bibinfo {author} {\bibfnamefont {Frederik}\
  \bibnamefont {Thiele}}, \bibinfo {author} {\bibfnamefont {Felix~vom}\
  \bibnamefont {Bruch}}, \bibinfo {author} {\bibfnamefont {Victor}\
  \bibnamefont {Quiring}}, \bibinfo {author} {\bibfnamefont {Raimund}\
  \bibnamefont {Ricken}}, \bibinfo {author} {\bibfnamefont {Harald}\
  \bibnamefont {Herrmann}}, \bibinfo {author} {\bibfnamefont {Christof}\
  \bibnamefont {Eigner}}, \bibinfo {author} {\bibfnamefont {Christine}\
  \bibnamefont {Silberhorn}}, \ and\ \bibinfo {author} {\bibfnamefont {Tim~J.}\
  \bibnamefont {Bartley}},\ }\bibfield  {title} {\enquote {\bibinfo {title}
  {Cryogenic {Electro}-{Optic} {Polarisation} {Conversion} in {Titanium}
  in-diffused {Lithium} {Niobate} {Waveguides}},}\ }\href
  {http://arxiv.org/abs/2006.12078} {\bibfield  {journal} {\bibinfo  {journal}
  {arXiv:2006.12078 [physics, physics:quant-ph]}\ } (\bibinfo {year}
  {2020})}\BibitemShut {NoStop}%
\bibitem [{\citenamefont {Chakraborty}\ \emph {et~al.}(2020)\citenamefont
  {Chakraborty}, \citenamefont {Carolan}, \citenamefont {Clark}, \citenamefont
  {Bunandar}, \citenamefont {Notaros}, \citenamefont {Watts},\ and\
  \citenamefont {Englund}}]{chakraborty2020cryogenic}%
  \BibitemOpen
  \bibfield  {author} {\bibinfo {author} {\bibfnamefont {Uttara}\ \bibnamefont
  {Chakraborty}}, \bibinfo {author} {\bibfnamefont {Jacques}\ \bibnamefont
  {Carolan}}, \bibinfo {author} {\bibfnamefont {Genevieve}\ \bibnamefont
  {Clark}}, \bibinfo {author} {\bibfnamefont {Darius}\ \bibnamefont
  {Bunandar}}, \bibinfo {author} {\bibfnamefont {Jelena}\ \bibnamefont
  {Notaros}}, \bibinfo {author} {\bibfnamefont {Michael~R}\ \bibnamefont
  {Watts}}, \ and\ \bibinfo {author} {\bibfnamefont {Dirk~R}\ \bibnamefont
  {Englund}},\ }\bibfield  {title} {\enquote {\bibinfo {title} {Cryogenic
  operation of silicon photonic modulators based on dc kerr effect},}\ }\href
  {https://arxiv.org/abs/2007.03395} {\bibfield  {journal} {\bibinfo  {journal}
  {arXiv preprint arXiv:2007.03395}\ } (\bibinfo {year} {2020})}\BibitemShut
  {NoStop}%
\bibitem [{\citenamefont {Abel}\ \emph {et~al.}(2019)\citenamefont {Abel},
  \citenamefont {Eltes}, \citenamefont {Ortmann}, \citenamefont {Messner},
  \citenamefont {Castera} \emph {et~al.}}]{Abel2019}%
  \BibitemOpen
  \bibfield  {author} {\bibinfo {author} {\bibfnamefont {Stefan}\ \bibnamefont
  {Abel}}, \bibinfo {author} {\bibfnamefont {Felix}\ \bibnamefont {Eltes}},
  \bibinfo {author} {\bibfnamefont {J.~Elliott}\ \bibnamefont {Ortmann}},
  \bibinfo {author} {\bibfnamefont {Andreas}\ \bibnamefont {Messner}}, \bibinfo
  {author} {\bibfnamefont {Pau}\ \bibnamefont {Castera}},  \emph {et~al.},\
  }\bibfield  {title} {\enquote {\bibinfo {title} {Large pockels effect in
  micro- and nanostructured barium titanate integrated on silicon},}\ }\href
  {\doibase 10.1038/s41563-018-0208-0} {\bibfield  {journal} {\bibinfo
  {journal} {Nature Materials}\ }\textbf {\bibinfo {volume} {18}},\ \bibinfo
  {pages} {42–47} (\bibinfo {year} {2019})}\BibitemShut {NoStop}%
\bibitem [{\citenamefont {Eltes}\ \emph {et~al.}(2019)\citenamefont {Eltes},
  \citenamefont {Villarreal-Garcia}, \citenamefont {Caimi}, \citenamefont
  {Siegwart}, \citenamefont {Gentile} \emph {et~al.}}]{Abel_BTO_cryo}%
  \BibitemOpen
  \bibfield  {author} {\bibinfo {author} {\bibfnamefont {Felix}\ \bibnamefont
  {Eltes}}, \bibinfo {author} {\bibfnamefont {Gerardo~E.}\ \bibnamefont
  {Villarreal-Garcia}}, \bibinfo {author} {\bibfnamefont {Daniele}\
  \bibnamefont {Caimi}}, \bibinfo {author} {\bibfnamefont {Heinz}\ \bibnamefont
  {Siegwart}}, \bibinfo {author} {\bibfnamefont {Antonio~A.}\ \bibnamefont
  {Gentile}},  \emph {et~al.},\ }\bibfield  {title} {\enquote {\bibinfo {title}
  {An integrated cryogenic optical modulator},}\ }\href
  {http://arxiv.org/abs/1904.10902} {\bibfield  {journal} {\bibinfo  {journal}
  {arXiv:1904.10902 [physics]}\ } (\bibinfo {year} {2019})},\ \bibinfo {note}
  {arXiv: 1904.10902}\BibitemShut {NoStop}%
\bibitem [{\citenamefont {Braginski}(2019)}]{braginski2019}%
  \BibitemOpen
  \bibfield  {author} {\bibinfo {author} {\bibfnamefont {Alex~I.}\ \bibnamefont
  {Braginski}},\ }\bibfield  {title} {\enquote {\bibinfo {title}
  {Superconductor {Electronics}: {Status} and {Outlook}},}\ }\href {\doibase
  10.1007/s10948-018-4884-4} {\bibfield  {journal} {\bibinfo  {journal}
  {Journal of Superconductivity and Novel Magnetism}\ }\textbf {\bibinfo
  {volume} {32}},\ \bibinfo {pages} {23--44} (\bibinfo {year}
  {2019})}\BibitemShut {NoStop}%
\bibitem [{\citenamefont {Cheng}\ \emph {et~al.}(2018)\citenamefont {Cheng},
  \citenamefont {Bahadori}, \citenamefont {Glick}, \citenamefont {Rumley},\
  and\ \citenamefont {Bergman}}]{Cheng2018}%
  \BibitemOpen
  \bibfield  {author} {\bibinfo {author} {\bibfnamefont {Qixiang}\ \bibnamefont
  {Cheng}}, \bibinfo {author} {\bibfnamefont {Meisam}\ \bibnamefont
  {Bahadori}}, \bibinfo {author} {\bibfnamefont {Madeleine}\ \bibnamefont
  {Glick}}, \bibinfo {author} {\bibfnamefont {S\'{e}bastien}\ \bibnamefont
  {Rumley}}, \ and\ \bibinfo {author} {\bibfnamefont {Keren}\ \bibnamefont
  {Bergman}},\ }\bibfield  {title} {\enquote {\bibinfo {title} {Recent advances
  in optical technologies for data centers: a review},}\ }\href {\doibase
  10.1364/OPTICA.5.001354} {\bibfield  {journal} {\bibinfo  {journal} {Optica}\
  }\textbf {\bibinfo {volume} {5}},\ \bibinfo {pages} {1354--1370} (\bibinfo
  {year} {2018})}\BibitemShut {NoStop}%
\bibitem [{\citenamefont {Thomson}\ \emph {et~al.}(2016)\citenamefont
  {Thomson}, \citenamefont {Zilkie}, \citenamefont {Bowers}, \citenamefont
  {Komljenovic}, \citenamefont {Reed} \emph {et~al.}}]{Thomson2016}%
  \BibitemOpen
  \bibfield  {author} {\bibinfo {author} {\bibfnamefont {David}\ \bibnamefont
  {Thomson}}, \bibinfo {author} {\bibfnamefont {Aaron}\ \bibnamefont {Zilkie}},
  \bibinfo {author} {\bibfnamefont {John~E}\ \bibnamefont {Bowers}}, \bibinfo
  {author} {\bibfnamefont {Tin}\ \bibnamefont {Komljenovic}}, \bibinfo {author}
  {\bibfnamefont {Graham~T}\ \bibnamefont {Reed}},  \emph {et~al.},\ }\bibfield
   {title} {\enquote {\bibinfo {title} {Roadmap on silicon photonics},}\ }\href
  {\doibase 10.1088/2040-8978/18/7/073003} {\bibfield  {journal} {\bibinfo
  {journal} {Journal of Optics}\ }\textbf {\bibinfo {volume} {18}},\ \bibinfo
  {pages} {073003} (\bibinfo {year} {2016})}\BibitemShut {NoStop}%
\bibitem [{\citenamefont {Blumenthal}\ \emph {et~al.}(2000)\citenamefont
  {Blumenthal}, \citenamefont {Olsson}, \citenamefont {Rossi}, \citenamefont
  {Dimmick}, \citenamefont {Rau} \emph {et~al.}}]{blumenthal2000all}%
  \BibitemOpen
  \bibfield  {author} {\bibinfo {author} {\bibfnamefont {Daniel~J}\
  \bibnamefont {Blumenthal}}, \bibinfo {author} {\bibfnamefont {Bengt-Erik}\
  \bibnamefont {Olsson}}, \bibinfo {author} {\bibfnamefont {Giammarco}\
  \bibnamefont {Rossi}}, \bibinfo {author} {\bibfnamefont {Timothy~E}\
  \bibnamefont {Dimmick}}, \bibinfo {author} {\bibfnamefont {Lavanya}\
  \bibnamefont {Rau}},  \emph {et~al.},\ }\bibfield  {title} {\enquote
  {\bibinfo {title} {All-optical label swapping networks and technologies},}\
  }\href {https://www.osapublishing.org/jlt/abstract.cfm?uri=jlt-18-12-2058}
  {\bibfield  {journal} {\bibinfo  {journal} {Journal of Lightwave Technology}\
  }\textbf {\bibinfo {volume} {18}},\ \bibinfo {pages} {2058} (\bibinfo {year}
  {2000})}\BibitemShut {NoStop}%
\bibitem [{\citenamefont {Sun}\ \emph {et~al.}(2015)\citenamefont {Sun},
  \citenamefont {Wade}, \citenamefont {Lee}, \citenamefont {Orcutt},
  \citenamefont {Alloatti} \emph {et~al.}}]{sun2015single}%
  \BibitemOpen
  \bibfield  {author} {\bibinfo {author} {\bibfnamefont {Chen}\ \bibnamefont
  {Sun}}, \bibinfo {author} {\bibfnamefont {Mark~T}\ \bibnamefont {Wade}},
  \bibinfo {author} {\bibfnamefont {Yunsup}\ \bibnamefont {Lee}}, \bibinfo
  {author} {\bibfnamefont {Jason~S}\ \bibnamefont {Orcutt}}, \bibinfo {author}
  {\bibfnamefont {Luca}\ \bibnamefont {Alloatti}},  \emph {et~al.},\ }\bibfield
   {title} {\enquote {\bibinfo {title} {Single-chip microprocessor that
  communicates directly using light},}\ }\href
  {https://www.nature.com/articles/nature16454?proof=true%3Ca+href=} {\bibfield
   {journal} {\bibinfo  {journal} {Nature}\ }\textbf {\bibinfo {volume}
  {528}},\ \bibinfo {pages} {534--538} (\bibinfo {year} {2015})}\BibitemShut
  {NoStop}%
\bibitem [{\citenamefont {Miller}(2000)}]{miller2000optical}%
  \BibitemOpen
  \bibfield  {author} {\bibinfo {author} {\bibfnamefont {David~AB}\
  \bibnamefont {Miller}},\ }\bibfield  {title} {\enquote {\bibinfo {title}
  {Optical interconnects to silicon},}\ }\href
  {https://ieeexplore.ieee.org/abstract/document/902184} {\bibfield  {journal}
  {\bibinfo  {journal} {IEEE Journal of Selected Topics in Quantum
  Electronics}\ }\textbf {\bibinfo {volume} {6}},\ \bibinfo {pages}
  {1312--1317} (\bibinfo {year} {2000})}\BibitemShut {NoStop}%
\bibitem [{\citenamefont {Arute}\ \emph {et~al.}(2019)\citenamefont {Arute},
  \citenamefont {Arya}, \citenamefont {Babbush}, \citenamefont {Bacon},
  \citenamefont {Bardin} \emph {et~al.}}]{arute2019quantum}%
  \BibitemOpen
  \bibfield  {author} {\bibinfo {author} {\bibfnamefont {Frank}\ \bibnamefont
  {Arute}}, \bibinfo {author} {\bibfnamefont {Kunal}\ \bibnamefont {Arya}},
  \bibinfo {author} {\bibfnamefont {Ryan}\ \bibnamefont {Babbush}}, \bibinfo
  {author} {\bibfnamefont {Dave}\ \bibnamefont {Bacon}}, \bibinfo {author}
  {\bibfnamefont {Joseph~C.}\ \bibnamefont {Bardin}},  \emph {et~al.},\
  }\bibfield  {title} {\enquote {\bibinfo {title} {Quantum supremacy using a
  programmable superconducting processor},}\ }\href {\doibase
  10.1038/s41586-019-1666-5} {\bibfield  {journal} {\bibinfo  {journal}
  {Nature}\ }\textbf {\bibinfo {volume} {574}},\ \bibinfo {pages} {505--510}
  (\bibinfo {year} {2019})}\BibitemShut {NoStop}%
\bibitem [{\citenamefont {Krinner}\ \emph {et~al.}(2019)\citenamefont
  {Krinner}, \citenamefont {Storz}, \citenamefont {Kurpiers}, \citenamefont
  {Magnard}, \citenamefont {Heinsoo} \emph {et~al.}}]{krinner2019}%
  \BibitemOpen
  \bibfield  {author} {\bibinfo {author} {\bibfnamefont {Sebastian}\
  \bibnamefont {Krinner}}, \bibinfo {author} {\bibfnamefont {Simon}\
  \bibnamefont {Storz}}, \bibinfo {author} {\bibfnamefont {Philipp}\
  \bibnamefont {Kurpiers}}, \bibinfo {author} {\bibfnamefont {Paul}\
  \bibnamefont {Magnard}}, \bibinfo {author} {\bibfnamefont {Johannes}\
  \bibnamefont {Heinsoo}},  \emph {et~al.},\ }\bibfield  {title} {\enquote
  {\bibinfo {title} {Engineering cryogenic setups for 100-qubit scale
  superconducting circuit systems},}\ }\href {\doibase
  10.1140/epjqt/s40507-019-0072-0} {\bibfield  {journal} {\bibinfo  {journal}
  {EPJ Quantum Technology}\ }\textbf {\bibinfo {volume} {6}},\ \bibinfo {pages}
  {1--29} (\bibinfo {year} {2019})}\BibitemShut {NoStop}%
\bibitem [{\citenamefont {Yamamoto}\ \emph {et~al.}(2008)\citenamefont
  {Yamamoto}, \citenamefont {Inomata}, \citenamefont {Watanabe}, \citenamefont
  {Matsuba}, \citenamefont {Miyazaki} \emph {et~al.}}]{yamamoto2008flux}%
  \BibitemOpen
  \bibfield  {author} {\bibinfo {author} {\bibfnamefont {Tsuyoshi}\
  \bibnamefont {Yamamoto}}, \bibinfo {author} {\bibfnamefont {K}~\bibnamefont
  {Inomata}}, \bibinfo {author} {\bibfnamefont {M}~\bibnamefont {Watanabe}},
  \bibinfo {author} {\bibfnamefont {K}~\bibnamefont {Matsuba}}, \bibinfo
  {author} {\bibfnamefont {T}~\bibnamefont {Miyazaki}},  \emph {et~al.},\
  }\bibfield  {title} {\enquote {\bibinfo {title} {Flux-driven josephson
  parametric amplifier},}\ }\href
  {https://aip.scitation.org/doi/full/10.1063/1.2964182} {\bibfield  {journal}
  {\bibinfo  {journal} {Applied Physics Letters}\ }\textbf {\bibinfo {volume}
  {93}},\ \bibinfo {pages} {042510} (\bibinfo {year} {2008})}\BibitemShut
  {NoStop}%
\bibitem [{\citenamefont {Macklin}\ \emph {et~al.}(2015)\citenamefont
  {Macklin}, \citenamefont {O’Brien}, \citenamefont {Hover}, \citenamefont
  {Schwartz}, \citenamefont {Bolkhovsky} \emph {et~al.}}]{macklin2015near}%
  \BibitemOpen
  \bibfield  {author} {\bibinfo {author} {\bibfnamefont {Chris}\ \bibnamefont
  {Macklin}}, \bibinfo {author} {\bibfnamefont {K}~\bibnamefont {O’Brien}},
  \bibinfo {author} {\bibfnamefont {D}~\bibnamefont {Hover}}, \bibinfo {author}
  {\bibfnamefont {ME}~\bibnamefont {Schwartz}}, \bibinfo {author}
  {\bibfnamefont {V}~\bibnamefont {Bolkhovsky}},  \emph {et~al.},\ }\bibfield
  {title} {\enquote {\bibinfo {title} {A near--quantum-limited josephson
  traveling-wave parametric amplifier},}\ }\href
  {https://science.sciencemag.org/content/350/6258/307} {\bibfield  {journal}
  {\bibinfo  {journal} {Science}\ }\textbf {\bibinfo {volume} {350}},\ \bibinfo
  {pages} {307--310} (\bibinfo {year} {2015})}\BibitemShut {NoStop}%
\bibitem [{\citenamefont {Siddiqi}\ \emph {et~al.}(2004)\citenamefont
  {Siddiqi}, \citenamefont {Vijay}, \citenamefont {Pierre}, \citenamefont
  {Wilson}, \citenamefont {Metcalfe} \emph {et~al.}}]{siddiqi2004rf}%
  \BibitemOpen
  \bibfield  {author} {\bibinfo {author} {\bibfnamefont {Irfan}\ \bibnamefont
  {Siddiqi}}, \bibinfo {author} {\bibfnamefont {R}~\bibnamefont {Vijay}},
  \bibinfo {author} {\bibfnamefont {F}~\bibnamefont {Pierre}}, \bibinfo
  {author} {\bibfnamefont {CM}~\bibnamefont {Wilson}}, \bibinfo {author}
  {\bibfnamefont {M}~\bibnamefont {Metcalfe}},  \emph {et~al.},\ }\bibfield
  {title} {\enquote {\bibinfo {title} {Rf-driven josephson bifurcation
  amplifier for quantum measurement},}\ }\href
  {https://journals.aps.org/prl/abstract/10.1103/PhysRevLett.93.207002}
  {\bibfield  {journal} {\bibinfo  {journal} {Physical review letters}\
  }\textbf {\bibinfo {volume} {93}},\ \bibinfo {pages} {207002} (\bibinfo
  {year} {2004})}\BibitemShut {NoStop}%
\bibitem [{\citenamefont {Castellanos-Beltran}\ \emph
  {et~al.}(2008)\citenamefont {Castellanos-Beltran}, \citenamefont {Irwin},
  \citenamefont {Hilton}, \citenamefont {Vale},\ and\ \citenamefont
  {Lehnert}}]{castellanos2008amplification}%
  \BibitemOpen
  \bibfield  {author} {\bibinfo {author} {\bibfnamefont {MA}~\bibnamefont
  {Castellanos-Beltran}}, \bibinfo {author} {\bibfnamefont {KD}~\bibnamefont
  {Irwin}}, \bibinfo {author} {\bibfnamefont {GC}~\bibnamefont {Hilton}},
  \bibinfo {author} {\bibfnamefont {LR}~\bibnamefont {Vale}}, \ and\ \bibinfo
  {author} {\bibfnamefont {KW}~\bibnamefont {Lehnert}},\ }\bibfield  {title}
  {\enquote {\bibinfo {title} {Amplification and squeezing of quantum noise
  with a tunable josephson metamaterial},}\ }\href
  {https://www.nature.com/articles/nphys1090} {\bibfield  {journal} {\bibinfo
  {journal} {Nature Physics}\ }\textbf {\bibinfo {volume} {4}},\ \bibinfo
  {pages} {929--931} (\bibinfo {year} {2008})}\BibitemShut {NoStop}%
\bibitem [{\citenamefont {Lauk}\ \emph {et~al.}(2020)\citenamefont {Lauk},
  \citenamefont {Sinclair}, \citenamefont {Barzanjeh}, \citenamefont {Covey},
  \citenamefont {Saffman} \emph {et~al.}}]{lauk2020perspectives}%
  \BibitemOpen
  \bibfield  {author} {\bibinfo {author} {\bibfnamefont {Nikolai}\ \bibnamefont
  {Lauk}}, \bibinfo {author} {\bibfnamefont {Neil}\ \bibnamefont {Sinclair}},
  \bibinfo {author} {\bibfnamefont {Shabir}\ \bibnamefont {Barzanjeh}},
  \bibinfo {author} {\bibfnamefont {Jacob~P}\ \bibnamefont {Covey}}, \bibinfo
  {author} {\bibfnamefont {Mark}\ \bibnamefont {Saffman}},  \emph {et~al.},\
  }\bibfield  {title} {\enquote {\bibinfo {title} {Perspectives on quantum
  transduction},}\ }\href
  {https://iopscience.iop.org/article/10.1088/2058-9565/ab788a} {\bibfield
  {journal} {\bibinfo  {journal} {Quantum Science and Technology}\ }\textbf
  {\bibinfo {volume} {5}},\ \bibinfo {pages} {020501} (\bibinfo {year}
  {2020})}\BibitemShut {NoStop}%
\bibitem [{\citenamefont {Jiang}\ \emph
  {et~al.}(2019{\natexlab{a}})\citenamefont {Jiang}, \citenamefont {Patel},
  \citenamefont {Mayor}, \citenamefont {McKenna}, \citenamefont
  {Arrangoiz-Arriola} \emph {et~al.}}]{jiang2019lithium}%
  \BibitemOpen
  \bibfield  {author} {\bibinfo {author} {\bibfnamefont {Wentao}\ \bibnamefont
  {Jiang}}, \bibinfo {author} {\bibfnamefont {Rishi~N}\ \bibnamefont {Patel}},
  \bibinfo {author} {\bibfnamefont {Felix~M}\ \bibnamefont {Mayor}}, \bibinfo
  {author} {\bibfnamefont {Timothy~P}\ \bibnamefont {McKenna}}, \bibinfo
  {author} {\bibfnamefont {Patricio}\ \bibnamefont {Arrangoiz-Arriola}},  \emph
  {et~al.},\ }\bibfield  {title} {\enquote {\bibinfo {title} {Lithium niobate
  piezo-optomechanical crystals},}\ }\href
  {https://www.osapublishing.org/optica/abstract.cfm?uri=optica-6-7-845}
  {\bibfield  {journal} {\bibinfo  {journal} {Optica}\ }\textbf {\bibinfo
  {volume} {6}},\ \bibinfo {pages} {845--853} (\bibinfo {year}
  {2019}{\natexlab{a}})}\BibitemShut {NoStop}%
\bibitem [{\citenamefont {Jiang}\ \emph
  {et~al.}(2019{\natexlab{b}})\citenamefont {Jiang}, \citenamefont {Sarabalis},
  \citenamefont {Dahmani}, \citenamefont {Patel}, \citenamefont {Mayor} \emph
  {et~al.}}]{jiang2019efficient}%
  \BibitemOpen
  \bibfield  {author} {\bibinfo {author} {\bibfnamefont {Wentao}\ \bibnamefont
  {Jiang}}, \bibinfo {author} {\bibfnamefont {Christopher~J}\ \bibnamefont
  {Sarabalis}}, \bibinfo {author} {\bibfnamefont {Yanni~D}\ \bibnamefont
  {Dahmani}}, \bibinfo {author} {\bibfnamefont {Rishi~N}\ \bibnamefont
  {Patel}}, \bibinfo {author} {\bibfnamefont {Felix~M}\ \bibnamefont {Mayor}},
  \emph {et~al.},\ }\bibfield  {title} {\enquote {\bibinfo {title} {Efficient
  bidirectional piezo-optomechanical transduction between microwave and optical
  frequency},}\ }\href {https://arxiv.org/abs/1909.04627} {\bibfield  {journal}
  {\bibinfo  {journal} {arXiv preprint arXiv:1909.04627}\ } (\bibinfo {year}
  {2019}{\natexlab{b}})}\BibitemShut {NoStop}%
\bibitem [{\citenamefont {Bartholomew}\ \emph {et~al.}(2019)\citenamefont
  {Bartholomew}, \citenamefont {Rochman}, \citenamefont {Xie}, \citenamefont
  {Kindem}, \citenamefont {Ruskuc} \emph {et~al.}}]{bartholomew2019chip}%
  \BibitemOpen
  \bibfield  {author} {\bibinfo {author} {\bibfnamefont {John~G}\ \bibnamefont
  {Bartholomew}}, \bibinfo {author} {\bibfnamefont {Jake}\ \bibnamefont
  {Rochman}}, \bibinfo {author} {\bibfnamefont {Tian}\ \bibnamefont {Xie}},
  \bibinfo {author} {\bibfnamefont {Jonathan~M}\ \bibnamefont {Kindem}},
  \bibinfo {author} {\bibfnamefont {Andrei}\ \bibnamefont {Ruskuc}},  \emph
  {et~al.},\ }\bibfield  {title} {\enquote {\bibinfo {title} {On-chip coherent
  microwave-to-optical transduction mediated by ytterbium in yvo $ \_4$},}\
  }\href {https://arxiv.org/abs/1912.03671} {\bibfield  {journal} {\bibinfo
  {journal} {arXiv preprint arXiv:1912.03671}\ } (\bibinfo {year}
  {2019})}\BibitemShut {NoStop}%
\bibitem [{\citenamefont {Higginbotham}\ \emph {et~al.}(2018)\citenamefont
  {Higginbotham}, \citenamefont {Burns}, \citenamefont {Urmey}, \citenamefont
  {Peterson}, \citenamefont {Kampel} \emph {et~al.}}]{Higginbotham2018}%
  \BibitemOpen
  \bibfield  {author} {\bibinfo {author} {\bibfnamefont {A.~P.}\ \bibnamefont
  {Higginbotham}}, \bibinfo {author} {\bibfnamefont {P.~S.}\ \bibnamefont
  {Burns}}, \bibinfo {author} {\bibfnamefont {M.~D.}\ \bibnamefont {Urmey}},
  \bibinfo {author} {\bibfnamefont {R.~W.}\ \bibnamefont {Peterson}}, \bibinfo
  {author} {\bibfnamefont {N.~S.}\ \bibnamefont {Kampel}},  \emph {et~al.},\
  }\bibfield  {title} {\enquote {\bibinfo {title} {Harnessing electro-optic
  correlations in an efficient mechanical converter},}\ }\href {\doibase
  10.1038/s41567-018-0210-0} {\bibfield  {journal} {\bibinfo  {journal} {Nature
  Physics}\ }\textbf {\bibinfo {volume} {14}},\ \bibinfo {pages} {1038–1042}
  (\bibinfo {year} {2018})}\BibitemShut {NoStop}%
\bibitem [{\citenamefont {Forsch}\ \emph {et~al.}(2020)\citenamefont {Forsch},
  \citenamefont {Stockill}, \citenamefont {Wallucks}, \citenamefont
  {Marinkovi{\'c}}, \citenamefont {G{\"a}rtner} \emph
  {et~al.}}]{forsch2020microwave}%
  \BibitemOpen
  \bibfield  {author} {\bibinfo {author} {\bibfnamefont {Moritz}\ \bibnamefont
  {Forsch}}, \bibinfo {author} {\bibfnamefont {Robert}\ \bibnamefont
  {Stockill}}, \bibinfo {author} {\bibfnamefont {Andreas}\ \bibnamefont
  {Wallucks}}, \bibinfo {author} {\bibfnamefont {Igor}\ \bibnamefont
  {Marinkovi{\'c}}}, \bibinfo {author} {\bibfnamefont {Claus}\ \bibnamefont
  {G{\"a}rtner}},  \emph {et~al.},\ }\bibfield  {title} {\enquote {\bibinfo
  {title} {Microwave-to-optics conversion using a mechanical oscillator in its
  quantum ground state},}\ }\href
  {https://www.nature.com/articles/s41567-019-0673-7} {\bibfield  {journal}
  {\bibinfo  {journal} {Nature Physics}\ }\textbf {\bibinfo {volume} {16}},\
  \bibinfo {pages} {69--74} (\bibinfo {year} {2020})}\BibitemShut {NoStop}%
\bibitem [{\citenamefont {Arnold}\ \emph {et~al.}(2020)\citenamefont {Arnold},
  \citenamefont {Wulf}, \citenamefont {Barzanjeh}, \citenamefont {Redchenko},
  \citenamefont {Rueda} \emph {et~al.}}]{arnold2020converting}%
  \BibitemOpen
  \bibfield  {author} {\bibinfo {author} {\bibfnamefont {G}~\bibnamefont
  {Arnold}}, \bibinfo {author} {\bibfnamefont {M}~\bibnamefont {Wulf}},
  \bibinfo {author} {\bibfnamefont {S}~\bibnamefont {Barzanjeh}}, \bibinfo
  {author} {\bibfnamefont {ES}~\bibnamefont {Redchenko}}, \bibinfo {author}
  {\bibfnamefont {A}~\bibnamefont {Rueda}},  \emph {et~al.},\ }\bibfield
  {title} {\enquote {\bibinfo {title} {Converting microwave and telecom photons
  with a silicon photonic nanomechanical interface},}\ }\href
  {https://arxiv.org/abs/2002.11628} {\bibfield  {journal} {\bibinfo  {journal}
  {arXiv preprint arXiv:2002.11628}\ } (\bibinfo {year} {2020})}\BibitemShut
  {NoStop}%
\bibitem [{\citenamefont {Andrews}\ \emph {et~al.}(2014)\citenamefont
  {Andrews}, \citenamefont {Peterson}, \citenamefont {Purdy}, \citenamefont
  {Cicak}, \citenamefont {Simmonds} \emph {et~al.}}]{andrews2014bidirectional}%
  \BibitemOpen
  \bibfield  {author} {\bibinfo {author} {\bibfnamefont {Reed~W}\ \bibnamefont
  {Andrews}}, \bibinfo {author} {\bibfnamefont {Robert~W}\ \bibnamefont
  {Peterson}}, \bibinfo {author} {\bibfnamefont {Tom~P}\ \bibnamefont {Purdy}},
  \bibinfo {author} {\bibfnamefont {Katarina}\ \bibnamefont {Cicak}}, \bibinfo
  {author} {\bibfnamefont {Raymond~W}\ \bibnamefont {Simmonds}},  \emph
  {et~al.},\ }\bibfield  {title} {\enquote {\bibinfo {title} {Bidirectional and
  efficient conversion between microwave and optical light},}\ }\href
  {https://www.nature.com/articles/nphys2911} {\bibfield  {journal} {\bibinfo
  {journal} {Nature Physics}\ }\textbf {\bibinfo {volume} {10}},\ \bibinfo
  {pages} {321--326} (\bibinfo {year} {2014})}\BibitemShut {NoStop}%
\bibitem [{\citenamefont {Chu}\ and\ \citenamefont
  {Gr{\"o}blacher}(2020)}]{chu2020perspective}%
  \BibitemOpen
  \bibfield  {author} {\bibinfo {author} {\bibfnamefont {Yiwen}\ \bibnamefont
  {Chu}}\ and\ \bibinfo {author} {\bibfnamefont {Simon}\ \bibnamefont
  {Gr{\"o}blacher}},\ }\bibfield  {title} {\enquote {\bibinfo {title} {A
  perspective on hybrid quantum opto-and electromechanical systems},}\ }\href
  {https://arxiv.org/abs/2007.03360} {\bibfield  {journal} {\bibinfo  {journal}
  {arXiv preprint arXiv:2007.03360}\ } (\bibinfo {year} {2020})}\BibitemShut
  {NoStop}%
\bibitem [{\citenamefont {Tsang}(2010)}]{tsang2010}%
  \BibitemOpen
  \bibfield  {author} {\bibinfo {author} {\bibfnamefont {Mankei}\ \bibnamefont
  {Tsang}},\ }\bibfield  {title} {\enquote {\bibinfo {title} {Cavity quantum
  electro-optics},}\ }\href {\doibase 10.1103/PhysRevA.81.063837} {\bibfield
  {journal} {\bibinfo  {journal} {Physical Review A}\ }\textbf {\bibinfo
  {volume} {81}},\ \bibinfo {pages} {063837} (\bibinfo {year}
  {2010})}\BibitemShut {NoStop}%
\bibitem [{\citenamefont {Rueda}\ \emph {et~al.}(2016)\citenamefont {Rueda},
  \citenamefont {Sedlmeir}, \citenamefont {Collodo}, \citenamefont {Vogl},
  \citenamefont {Stiller} \emph {et~al.}}]{rueda2016efficient}%
  \BibitemOpen
  \bibfield  {author} {\bibinfo {author} {\bibfnamefont {Alfredo}\ \bibnamefont
  {Rueda}}, \bibinfo {author} {\bibfnamefont {Florian}\ \bibnamefont
  {Sedlmeir}}, \bibinfo {author} {\bibfnamefont {Michele~C}\ \bibnamefont
  {Collodo}}, \bibinfo {author} {\bibfnamefont {Ulrich}\ \bibnamefont {Vogl}},
  \bibinfo {author} {\bibfnamefont {Birgit}\ \bibnamefont {Stiller}},  \emph
  {et~al.},\ }\bibfield  {title} {\enquote {\bibinfo {title} {Efficient
  microwave to optical photon conversion: an electro-optical realization},}\
  }\href {https://www.osapublishing.org/optica/abstract.cfm?uri=optica-3-6-597}
  {\bibfield  {journal} {\bibinfo  {journal} {Optica}\ }\textbf {\bibinfo
  {volume} {3}},\ \bibinfo {pages} {597--604} (\bibinfo {year}
  {2016})}\BibitemShut {NoStop}%
\bibitem [{\citenamefont {Rueda}\ \emph {et~al.}(2019)\citenamefont {Rueda},
  \citenamefont {Sedlmeir}, \citenamefont {Kumari}, \citenamefont {Leuchs},\
  and\ \citenamefont {Schwefel}}]{Rueda2019}%
  \BibitemOpen
  \bibfield  {author} {\bibinfo {author} {\bibfnamefont {Alfredo}\ \bibnamefont
  {Rueda}}, \bibinfo {author} {\bibfnamefont {Florian}\ \bibnamefont
  {Sedlmeir}}, \bibinfo {author} {\bibfnamefont {Madhuri}\ \bibnamefont
  {Kumari}}, \bibinfo {author} {\bibfnamefont {Gerd}\ \bibnamefont {Leuchs}}, \
  and\ \bibinfo {author} {\bibfnamefont {Harald G.~L.}\ \bibnamefont
  {Schwefel}},\ }\bibfield  {title} {\enquote {\bibinfo {title} {Resonant
  electro-optic frequency comb},}\ }\href {\doibase 10.1038/s41586-019-1110-x}
  {\bibfield  {journal} {\bibinfo  {journal} {Nature}\ }\textbf {\bibinfo
  {volume} {568}},\ \bibinfo {pages} {378–381} (\bibinfo {year}
  {2019})}\BibitemShut {NoStop}%
\bibitem [{\citenamefont {Hease}\ \emph {et~al.}(2020)\citenamefont {Hease},
  \citenamefont {Rueda}, \citenamefont {Sahu}, \citenamefont {Wulf},
  \citenamefont {Arnold}, \citenamefont {Schwefel},\ and\ \citenamefont
  {Fink}}]{hease2020}%
  \BibitemOpen
  \bibfield  {author} {\bibinfo {author} {\bibfnamefont {William}\ \bibnamefont
  {Hease}}, \bibinfo {author} {\bibfnamefont {Alfredo}\ \bibnamefont {Rueda}},
  \bibinfo {author} {\bibfnamefont {Rishabh}\ \bibnamefont {Sahu}}, \bibinfo
  {author} {\bibfnamefont {Matthias}\ \bibnamefont {Wulf}}, \bibinfo {author}
  {\bibfnamefont {Georg}\ \bibnamefont {Arnold}}, \bibinfo {author}
  {\bibfnamefont {Harald G.~L.}\ \bibnamefont {Schwefel}}, \ and\ \bibinfo
  {author} {\bibfnamefont {Johannes~M.}\ \bibnamefont {Fink}},\ }\bibfield
  {title} {\enquote {\bibinfo {title} {Cavity quantum electro-optics:
  {Microwave}-telecom conversion in the quantum ground state},}\ }\href
  {http://arxiv.org/abs/2005.12763} {\bibfield  {journal} {\bibinfo  {journal}
  {arXiv:2005.12763 [physics, physics:quant-ph]}\ } (\bibinfo {year}
  {2020})}\BibitemShut {NoStop}%
\bibitem [{\citenamefont {Fan}\ \emph {et~al.}(2018)\citenamefont {Fan},
  \citenamefont {Zou}, \citenamefont {Cheng}, \citenamefont {Guo},
  \citenamefont {Han} \emph {et~al.}}]{Faneaar2018}%
  \BibitemOpen
  \bibfield  {author} {\bibinfo {author} {\bibfnamefont {Linran}\ \bibnamefont
  {Fan}}, \bibinfo {author} {\bibfnamefont {Chang-Ling}\ \bibnamefont {Zou}},
  \bibinfo {author} {\bibfnamefont {Risheng}\ \bibnamefont {Cheng}}, \bibinfo
  {author} {\bibfnamefont {Xiang}\ \bibnamefont {Guo}}, \bibinfo {author}
  {\bibfnamefont {Xu}~\bibnamefont {Han}},  \emph {et~al.},\ }\bibfield
  {title} {\enquote {\bibinfo {title} {Superconducting cavity electro-optics: A
  platform for coherent photon conversion between superconducting and photonic
  circuits},}\ }\href {\doibase 10.1126/sciadv.aar4994} {\bibfield  {journal}
  {\bibinfo  {journal} {Science Advances}\ }\textbf {\bibinfo {volume} {4}}
  (\bibinfo {year} {2018}),\ 10.1126/sciadv.aar4994}\BibitemShut {NoStop}%
\bibitem [{\citenamefont {McKenna}\ \emph {et~al.}(2020)\citenamefont
  {McKenna}, \citenamefont {Witmer}, \citenamefont {Patel}, \citenamefont
  {Jiang}, \citenamefont {Van~Laer}, \citenamefont {Arrangoiz-Arriola},
  \citenamefont {Wollack}, \citenamefont {Herrmann},\ and\ \citenamefont
  {Safavi-Naeini}}]{mckenna2020}%
  \BibitemOpen
  \bibfield  {author} {\bibinfo {author} {\bibfnamefont {Timothy~P.}\
  \bibnamefont {McKenna}}, \bibinfo {author} {\bibfnamefont {Jeremy~D.}\
  \bibnamefont {Witmer}}, \bibinfo {author} {\bibfnamefont {Rishi~N.}\
  \bibnamefont {Patel}}, \bibinfo {author} {\bibfnamefont {Wentao}\
  \bibnamefont {Jiang}}, \bibinfo {author} {\bibfnamefont {Rapha\"el}\
  \bibnamefont {Van~Laer}}, \bibinfo {author} {\bibfnamefont {Patricio}\
  \bibnamefont {Arrangoiz-Arriola}}, \bibinfo {author} {\bibfnamefont
  {E.~Alex}\ \bibnamefont {Wollack}}, \bibinfo {author} {\bibfnamefont
  {Jason~F.}\ \bibnamefont {Herrmann}}, \ and\ \bibinfo {author} {\bibfnamefont
  {Amir~H.}\ \bibnamefont {Safavi-Naeini}},\ }\bibfield  {title} {\enquote
  {\bibinfo {title} {Cryogenic microwave-to-optical conversion using a
  triply-resonant lithium niobate on sapphire transducer},}\ }\href
  {http://arxiv.org/abs/2005.00897} {\bibfield  {journal} {\bibinfo  {journal}
  {arXiv:2005.00897 [physics, physics:quant-ph]}\ } (\bibinfo {year}
  {2020})}\BibitemShut {NoStop}%
\bibitem [{\citenamefont {Holzgrafe}\ \emph {et~al.}(2020)\citenamefont
  {Holzgrafe}, \citenamefont {Sinclair}, \citenamefont {Zhu}, \citenamefont
  {Shams-Ansari}, \citenamefont {Colangelo}, \citenamefont {Hu}, \citenamefont
  {Zhang}, \citenamefont {Berggren},\ and\ \citenamefont
  {Lon\v{c}ar}}]{holzgrafe2020}%
  \BibitemOpen
  \bibfield  {author} {\bibinfo {author} {\bibfnamefont {Jeffrey}\ \bibnamefont
  {Holzgrafe}}, \bibinfo {author} {\bibfnamefont {Neil}\ \bibnamefont
  {Sinclair}}, \bibinfo {author} {\bibfnamefont {Di}~\bibnamefont {Zhu}},
  \bibinfo {author} {\bibfnamefont {Amirhassan}\ \bibnamefont {Shams-Ansari}},
  \bibinfo {author} {\bibfnamefont {Marco}\ \bibnamefont {Colangelo}}, \bibinfo
  {author} {\bibfnamefont {Yaowen}\ \bibnamefont {Hu}}, \bibinfo {author}
  {\bibfnamefont {Mian}\ \bibnamefont {Zhang}}, \bibinfo {author}
  {\bibfnamefont {Karl~K.}\ \bibnamefont {Berggren}}, \ and\ \bibinfo {author}
  {\bibfnamefont {Marko}\ \bibnamefont {Lon\v{c}ar}},\ }\bibfield  {title}
  {\enquote {\bibinfo {title} {Cavity electro-optics in thin-film lithium
  niobate for efficient microwave-to-optical transduction},}\ }\href
  {http://arxiv.org/abs/2005.00939} {\bibfield  {journal} {\bibinfo  {journal}
  {arXiv:2005.00939 [physics, physics:quant-ph]}\ } (\bibinfo {year}
  {2020})}\BibitemShut {NoStop}%
\bibitem [{\citenamefont {Clerk}\ \emph {et~al.}(2010)\citenamefont {Clerk},
  \citenamefont {Devoret}, \citenamefont {Girvin}, \citenamefont {Marquardt},\
  and\ \citenamefont {Schoelkopf}}]{clerk2010}%
  \BibitemOpen
  \bibfield  {author} {\bibinfo {author} {\bibfnamefont {A.~A.}\ \bibnamefont
  {Clerk}}, \bibinfo {author} {\bibfnamefont {M.~H.}\ \bibnamefont {Devoret}},
  \bibinfo {author} {\bibfnamefont {S.~M.}\ \bibnamefont {Girvin}}, \bibinfo
  {author} {\bibfnamefont {Florian}\ \bibnamefont {Marquardt}}, \ and\ \bibinfo
  {author} {\bibfnamefont {R.~J.}\ \bibnamefont {Schoelkopf}},\ }\bibfield
  {title} {\enquote {\bibinfo {title} {Introduction to quantum noise,
  measurement, and amplification},}\ }\href {\doibase
  10.1103/RevModPhys.82.1155} {\bibfield  {journal} {\bibinfo  {journal}
  {Reviews of Modern Physics}\ }\textbf {\bibinfo {volume} {82}},\ \bibinfo
  {pages} {1155--1208} (\bibinfo {year} {2010})}\BibitemShut {NoStop}%
\bibitem [{\citenamefont {Herzog}\ \emph {et~al.}(2008)\citenamefont {Herzog},
  \citenamefont {Poberaj},\ and\ \citenamefont {Günter}}]{Herzog2008}%
  \BibitemOpen
  \bibfield  {author} {\bibinfo {author} {\bibfnamefont {Christian}\
  \bibnamefont {Herzog}}, \bibinfo {author} {\bibfnamefont {Gorazd}\
  \bibnamefont {Poberaj}}, \ and\ \bibinfo {author} {\bibfnamefont {Peter}\
  \bibnamefont {Günter}},\ }\bibfield  {title} {\enquote {\bibinfo {title}
  {Electro-optic behavior of lithium niobate at cryogenic temperatures},}\
  }\href {\doibase 10.1016/j.optcom.2007.10.031} {\bibfield  {journal}
  {\bibinfo  {journal} {Optics Communications}\ }\textbf {\bibinfo {volume}
  {281}},\ \bibinfo {pages} {793–796} (\bibinfo {year} {2008})}\BibitemShut
  {NoStop}%
\bibitem [{\citenamefont {Morse}\ \emph {et~al.}(1994)\citenamefont {Morse},
  \citenamefont {McCammon}, \citenamefont {McConaghy}, \citenamefont
  {Masquelier}, \citenamefont {Garrett} \emph {et~al.}}]{Morse1994}%
  \BibitemOpen
  \bibfield  {author} {\bibinfo {author} {\bibfnamefont {Jeffrey~D.}\
  \bibnamefont {Morse}}, \bibinfo {author} {\bibfnamefont {Kent~George}\
  \bibnamefont {McCammon}}, \bibinfo {author} {\bibfnamefont {Charles~F.}\
  \bibnamefont {McConaghy}}, \bibinfo {author} {\bibfnamefont {Don~A.}\
  \bibnamefont {Masquelier}}, \bibinfo {author} {\bibfnamefont {Henry~E.}\
  \bibnamefont {Garrett}},  \emph {et~al.},\ }\bibfield  {title} {\enquote
  {\bibinfo {title} {{Characterization of lithium niobate electro-optic
  modulators at cryogenic temperatures}},}\ }in\ \href {\doibase
  10.1117/12.175011} {\emph {\bibinfo {booktitle} {Design, Simulation, and
  Fabrication of Optoelectronic Devices and Circuits}}},\ Vol.\ \bibinfo
  {volume} {2150},\ \bibinfo {editor} {edited by\ \bibinfo {editor}
  {\bibfnamefont {Mario~Nicola}\ \bibnamefont {Armenise}}},\ \bibinfo
  {organization} {International Society for Optics and Photonics}\ (\bibinfo
  {publisher} {SPIE},\ \bibinfo {year} {1994})\ pp.\ \bibinfo {pages}
  {283--291}\BibitemShut {NoStop}%
\bibitem [{\citenamefont {McConaghy}\ \emph {et~al.}(1996)\citenamefont
  {McConaghy}, \citenamefont {Lowry}, \citenamefont {Becker},\ and\
  \citenamefont {Kincaid}}]{McConaghy1996}%
  \BibitemOpen
  \bibfield  {author} {\bibinfo {author} {\bibfnamefont {C.}~\bibnamefont
  {McConaghy}}, \bibinfo {author} {\bibfnamefont {M.}~\bibnamefont {Lowry}},
  \bibinfo {author} {\bibfnamefont {R.A.}\ \bibnamefont {Becker}}, \ and\
  \bibinfo {author} {\bibfnamefont {B.E.}\ \bibnamefont {Kincaid}},\ }\bibfield
   {title} {\enquote {\bibinfo {title} {The performance of pigtailed annealed
  proton exchange linbo 3 modulators at cryogenic temperatures},}\ }\href
  {\doibase 10.1109/68.541556} {\bibfield  {journal} {\bibinfo  {journal} {IEEE
  Photonics Technology Letters}\ }\textbf {\bibinfo {volume} {8}},\ \bibinfo
  {pages} {1480–1482} (\bibinfo {year} {1996})}\BibitemShut {NoStop}%
\bibitem [{\citenamefont {Yoshida}\ \emph {et~al.}(1999)\citenamefont
  {Yoshida}, \citenamefont {Kanda},\ and\ \citenamefont
  {Kohjiro}}]{Yoshida1999}%
  \BibitemOpen
  \bibfield  {author} {\bibinfo {author} {\bibfnamefont {K.}~\bibnamefont
  {Yoshida}}, \bibinfo {author} {\bibfnamefont {Y.}~\bibnamefont {Kanda}}, \
  and\ \bibinfo {author} {\bibfnamefont {S.}~\bibnamefont {Kohjiro}},\
  }\bibfield  {title} {\enquote {\bibinfo {title} {A traveling-wave-type
  {LiNbO$_3$} optical modulator with superconducting electrodes},}\ }\href
  {\doibase 10.1109/22.775458} {\bibfield  {journal} {\bibinfo  {journal} {IEEE
  Transactions on Microwave Theory and Techniques}\ }\textbf {\bibinfo {volume}
  {47}},\ \bibinfo {pages} {1201–1205} (\bibinfo {year} {1999})}\BibitemShut
  {NoStop}%
\bibitem [{BER(2018)}]{BER2018}%
  \BibitemOpen
  \href
  {https://www.keysight.com/ch/de/assets/7018-06435/white-papers/5992-3524.pdf}
  {\bibfield  {journal} {\bibinfo  {journal} {Keysight. How to measure BER}\ }
  (\bibinfo {year} {2018})}\BibitemShut {NoStop}%
\bibitem [{\citenamefont {Teufel}\ \emph
  {et~al.}(2011{\natexlab{b}})\citenamefont {Teufel}, \citenamefont {Donner},
  \citenamefont {Li}, \citenamefont {Harlow}, \citenamefont {Allman} \emph
  {et~al.}}]{Teufel2011}%
  \BibitemOpen
  \bibfield  {author} {\bibinfo {author} {\bibfnamefont {J.~D.}\ \bibnamefont
  {Teufel}}, \bibinfo {author} {\bibfnamefont {T.}~\bibnamefont {Donner}},
  \bibinfo {author} {\bibfnamefont {Dale}\ \bibnamefont {Li}}, \bibinfo
  {author} {\bibfnamefont {J.~W.}\ \bibnamefont {Harlow}}, \bibinfo {author}
  {\bibfnamefont {M.~S.}\ \bibnamefont {Allman}},  \emph {et~al.},\ }\bibfield
  {title} {\enquote {\bibinfo {title} {Sideband cooling of micromechanical
  motion to the quantum ground state},}\ }\href {\doibase 10.1038/nature10261}
  {\bibfield  {journal} {\bibinfo  {journal} {Nature}\ }\textbf {\bibinfo
  {volume} {475}},\ \bibinfo {pages} {359–363} (\bibinfo {year}
  {2011}{\natexlab{b}})}\BibitemShut {NoStop}%
\bibitem [{\citenamefont {Wollman}\ \emph {et~al.}(2015)\citenamefont
  {Wollman}, \citenamefont {Lei}, \citenamefont {Weinstein}, \citenamefont
  {Suh}, \citenamefont {Kronwald} \emph {et~al.}}]{Wollman2015}%
  \BibitemOpen
  \bibfield  {author} {\bibinfo {author} {\bibfnamefont {E.~E.}\ \bibnamefont
  {Wollman}}, \bibinfo {author} {\bibfnamefont {C.~U.}\ \bibnamefont {Lei}},
  \bibinfo {author} {\bibfnamefont {A.~J.}\ \bibnamefont {Weinstein}}, \bibinfo
  {author} {\bibfnamefont {J.}~\bibnamefont {Suh}}, \bibinfo {author}
  {\bibfnamefont {A.}~\bibnamefont {Kronwald}},  \emph {et~al.},\ }\bibfield
  {title} {\enquote {\bibinfo {title} {Quantum squeezing of motion in a
  mechanical resonator},}\ }\href {\doibase 10.1126/science.aac5138} {\bibfield
   {journal} {\bibinfo  {journal} {Science}\ }\textbf {\bibinfo {volume}
  {349}},\ \bibinfo {pages} {952–955} (\bibinfo {year} {2015})}\BibitemShut
  {NoStop}%
\bibitem [{\citenamefont {Ockeloen-Korppi}\ \emph {et~al.}(2018)\citenamefont
  {Ockeloen-Korppi}, \citenamefont {Damsk{\"a}gg}, \citenamefont
  {Pirkkalainen}, \citenamefont {Asjad}, \citenamefont {Clerk} \emph
  {et~al.}}]{ockeloen2018stabilized}%
  \BibitemOpen
  \bibfield  {author} {\bibinfo {author} {\bibfnamefont {CF}~\bibnamefont
  {Ockeloen-Korppi}}, \bibinfo {author} {\bibfnamefont {E}~\bibnamefont
  {Damsk{\"a}gg}}, \bibinfo {author} {\bibfnamefont {J-M}\ \bibnamefont
  {Pirkkalainen}}, \bibinfo {author} {\bibfnamefont {M}~\bibnamefont {Asjad}},
  \bibinfo {author} {\bibfnamefont {AA}~\bibnamefont {Clerk}},  \emph
  {et~al.},\ }\bibfield  {title} {\enquote {\bibinfo {title} {Stabilized
  entanglement of massive mechanical oscillators},}\ }\href
  {https://www.nature.com/articles/s41586-018-0038-x} {\bibfield  {journal}
  {\bibinfo  {journal} {Nature}\ }\textbf {\bibinfo {volume} {556}},\ \bibinfo
  {pages} {478--482} (\bibinfo {year} {2018})}\BibitemShut {NoStop}%
\bibitem [{\citenamefont {Barzanjeh}\ \emph {et~al.}(2019)\citenamefont
  {Barzanjeh}, \citenamefont {Redchenko}, \citenamefont {Peruzzo},
  \citenamefont {Wulf}, \citenamefont {Lewis} \emph
  {et~al.}}]{barzanjeh2019stationary}%
  \BibitemOpen
  \bibfield  {author} {\bibinfo {author} {\bibfnamefont {Shabir}\ \bibnamefont
  {Barzanjeh}}, \bibinfo {author} {\bibfnamefont {ES}~\bibnamefont
  {Redchenko}}, \bibinfo {author} {\bibfnamefont {Matilda}\ \bibnamefont
  {Peruzzo}}, \bibinfo {author} {\bibfnamefont {Matthias}\ \bibnamefont
  {Wulf}}, \bibinfo {author} {\bibfnamefont {DP}~\bibnamefont {Lewis}},  \emph
  {et~al.},\ }\bibfield  {title} {\enquote {\bibinfo {title} {Stationary
  entangled radiation from micromechanical motion},}\ }\href
  {https://www.nature.com/articles/s41586-019-1320-2} {\bibfield  {journal}
  {\bibinfo  {journal} {Nature}\ }\textbf {\bibinfo {volume} {570}},\ \bibinfo
  {pages} {480--483} (\bibinfo {year} {2019})}\BibitemShut {NoStop}%
\bibitem [{\citenamefont {Bernier}\ \emph {et~al.}(2017)\citenamefont
  {Bernier}, \citenamefont {Tóth}, \citenamefont {Koottandavida},
  \citenamefont {Ioannou}, \citenamefont {Malz} \emph {et~al.}}]{Bernier2017}%
  \BibitemOpen
  \bibfield  {author} {\bibinfo {author} {\bibfnamefont {N.~R.}\ \bibnamefont
  {Bernier}}, \bibinfo {author} {\bibfnamefont {L.~D.}\ \bibnamefont {Tóth}},
  \bibinfo {author} {\bibfnamefont {A.}~\bibnamefont {Koottandavida}}, \bibinfo
  {author} {\bibfnamefont {M.~A.}\ \bibnamefont {Ioannou}}, \bibinfo {author}
  {\bibfnamefont {D.}~\bibnamefont {Malz}},  \emph {et~al.},\ }\bibfield
  {title} {\enquote {\bibinfo {title} {Nonreciprocal reconfigurable microwave
  optomechanical circuit},}\ }\href {\doibase 10.1038/s41467-017-00447-1}
  {\bibfield  {journal} {\bibinfo  {journal} {Nature Communications}\ }\textbf
  {\bibinfo {volume} {8}} (\bibinfo {year} {2017}),\
  10.1038/s41467-017-00447-1}\BibitemShut {NoStop}%
\bibitem [{\citenamefont {Tóth}\ \emph {et~al.}(2017)\citenamefont {Tóth},
  \citenamefont {Bernier}, \citenamefont {Nunnenkamp}, \citenamefont
  {Feofanov},\ and\ \citenamefont {Kippenberg}}]{Toth2017}%
  \BibitemOpen
  \bibfield  {author} {\bibinfo {author} {\bibfnamefont {L.~D.}\ \bibnamefont
  {Tóth}}, \bibinfo {author} {\bibfnamefont {N.~R.}\ \bibnamefont {Bernier}},
  \bibinfo {author} {\bibfnamefont {A.}~\bibnamefont {Nunnenkamp}}, \bibinfo
  {author} {\bibfnamefont {A.~K.}\ \bibnamefont {Feofanov}}, \ and\ \bibinfo
  {author} {\bibfnamefont {T.~J.}\ \bibnamefont {Kippenberg}},\ }\bibfield
  {title} {\enquote {\bibinfo {title} {A dissipative quantum reservoir for
  microwave light using a mechanical oscillator},}\ }\href {\doibase
  10.1038/nphys4121} {\bibfield  {journal} {\bibinfo  {journal} {Nature
  Physics}\ }\textbf {\bibinfo {volume} {13}},\ \bibinfo {pages} {787–793}
  (\bibinfo {year} {2017})}\BibitemShut {NoStop}%
\bibitem [{\citenamefont {Aspelmeyer}\ \emph {et~al.}(2014)\citenamefont
  {Aspelmeyer}, \citenamefont {Kippenberg},\ and\ \citenamefont
  {Marquardt}}]{RMP_optomechanics}%
  \BibitemOpen
  \bibfield  {author} {\bibinfo {author} {\bibfnamefont {Markus}\ \bibnamefont
  {Aspelmeyer}}, \bibinfo {author} {\bibfnamefont {Tobias~J}\ \bibnamefont
  {Kippenberg}}, \ and\ \bibinfo {author} {\bibfnamefont {Florian}\
  \bibnamefont {Marquardt}},\ }\bibfield  {title} {\enquote {\bibinfo {title}
  {Cavity optomechanics},}\ }\href
  {https://journals.aps.org/rmp/abstract/10.1103/RevModPhys.86.1391} {\bibfield
   {journal} {\bibinfo  {journal} {Reviews of Modern Physics}\ }\textbf
  {\bibinfo {volume} {86}},\ \bibinfo {pages} {1391} (\bibinfo {year}
  {2014})}\BibitemShut {NoStop}%
\bibitem [{\citenamefont {Bernier}(2018)}]{NathanPhD}%
  \BibitemOpen
  \bibfield  {author} {\bibinfo {author} {\bibfnamefont {Nathan~R}\
  \bibnamefont {Bernier}},\ }\emph {\bibinfo {title} {Multimode microwave
  circuit optomechanics as a platform to study coupled quantum harmonic
  oscillators}},\ \href@noop {} {\bibinfo {type} {doctoral dissertation}},\
  \bibinfo  {school} {EPFL} (\bibinfo {year} {2018}),\ \bibinfo {note} {section
  3.5.2}\BibitemShut {NoStop}%
\bibitem [{\citenamefont {Marquardt}\ \emph {et~al.}(2006)\citenamefont
  {Marquardt}, \citenamefont {Harris},\ and\ \citenamefont
  {Girvin}}]{marquardt2006dynamical}%
  \BibitemOpen
  \bibfield  {author} {\bibinfo {author} {\bibfnamefont {Florian}\ \bibnamefont
  {Marquardt}}, \bibinfo {author} {\bibfnamefont {JGE}\ \bibnamefont {Harris}},
  \ and\ \bibinfo {author} {\bibfnamefont {Steven~M}\ \bibnamefont {Girvin}},\
  }\bibfield  {title} {\enquote {\bibinfo {title} {Dynamical multistability
  induced by radiation pressure in high-finesse micromechanical optical
  cavities},}\ }\href
  {https://journals.aps.org/prl/abstract/10.1103/PhysRevLett.96.103901}
  {\bibfield  {journal} {\bibinfo  {journal} {Physical Review Letters}\
  }\textbf {\bibinfo {volume} {96}},\ \bibinfo {pages} {103901} (\bibinfo
  {year} {2006})}\BibitemShut {NoStop}%
\bibitem [{\citenamefont {Carmon}\ \emph {et~al.}(2005)\citenamefont {Carmon},
  \citenamefont {Rokhsari}, \citenamefont {Yang}, \citenamefont {Kippenberg},\
  and\ \citenamefont {Vahala}}]{carmon2005temporal}%
  \BibitemOpen
  \bibfield  {author} {\bibinfo {author} {\bibfnamefont {Tal}\ \bibnamefont
  {Carmon}}, \bibinfo {author} {\bibfnamefont {Hossein}\ \bibnamefont
  {Rokhsari}}, \bibinfo {author} {\bibfnamefont {Lan}\ \bibnamefont {Yang}},
  \bibinfo {author} {\bibfnamefont {Tobias~J}\ \bibnamefont {Kippenberg}}, \
  and\ \bibinfo {author} {\bibfnamefont {Kerry~J}\ \bibnamefont {Vahala}},\
  }\bibfield  {title} {\enquote {\bibinfo {title} {Temporal behavior of
  radiation-pressure-induced vibrations of an optical microcavity phonon
  mode},}\ }\href
  {https://journals.aps.org/prl/abstract/10.1103/PhysRevLett.94.223902}
  {\bibfield  {journal} {\bibinfo  {journal} {Physical Review Letters}\
  }\textbf {\bibinfo {volume} {94}},\ \bibinfo {pages} {223902} (\bibinfo
  {year} {2005})}\BibitemShut {NoStop}%
\bibitem [{\citenamefont {Cattiaux}\ \emph {et~al.}(2020)\citenamefont
  {Cattiaux}, \citenamefont {Zhou}, \citenamefont {Kumar}, \citenamefont
  {Golokolenov}, \citenamefont {Gazizulin} \emph
  {et~al.}}]{cattiaux2020beyond}%
  \BibitemOpen
  \bibfield  {author} {\bibinfo {author} {\bibfnamefont {D}~\bibnamefont
  {Cattiaux}}, \bibinfo {author} {\bibfnamefont {X}~\bibnamefont {Zhou}},
  \bibinfo {author} {\bibfnamefont {S}~\bibnamefont {Kumar}}, \bibinfo {author}
  {\bibfnamefont {I}~\bibnamefont {Golokolenov}}, \bibinfo {author}
  {\bibfnamefont {RR}~\bibnamefont {Gazizulin}},  \emph {et~al.},\ }\bibfield
  {title} {\enquote {\bibinfo {title} {Beyond linear coupling in microwave
  optomechanics},}\ }\href {https://arxiv.org/abs/2003.03176} {\bibfield
  {journal} {\bibinfo  {journal} {arXiv preprint arXiv:2003.03176}\ } (\bibinfo
  {year} {2020})}\BibitemShut {NoStop}%
\bibitem [{Note1()}]{Note1}%
  \BibitemOpen
  \bibinfo {note} {Commercial PMs with typical insertion loss $<2\protect
  \ensuremath {\protect \tmspace +\thinmuskip {.1667em}\protect \mathrm {dB}}$,
  ($>63\%$ transmission), at room temperature are already
  available.}\BibitemShut {Stop}%
\end{thebibliography}%

\end{document}